\documentclass[twocolumn,
groupedaddress,
superscriptaddress,
amsmath,amssymb,
aps,
pra,
]{revtex4-1}
\usepackage{amsmath}
\usepackage{subfigure}
\usepackage{graphicx}
\usepackage{hyperref}
\hypersetup{
	colorlinks = true, linkcolor = blue, anchorcolor = blue, urlcolor = blue, citecolor = blue
}

\begin{document}
	
	
\title{Robust and stable delay interferometers with application to $d$-dimensional time-frequency quantum key distribution}
	
	
\author{Nurul T. Islam}
\affiliation{Department of Physics and the Fitzpatrick Institute for Photonics, Duke University, Durham, North Carolina 27708, USA}
\email[]{nti3@duke.edu}
\author{Andr\'{e}s Aragoneses}
\affiliation{Department of Physics and the Fitzpatrick Institute for Photonics, Duke University, Durham, North Carolina 27708, USA}
\affiliation{Department of Physics and Astronomy, Carleton College, Northfield, Minnesota 55057, USA}
\author{A. Lezama}
\affiliation{Department of Physics and the Fitzpatrick Institute for Photonics, Duke University, Durham, North Carolina 27708, USA}
\affiliation{Instituto de F\'{i}sica, Universidad de la Rep\'{u}blica. Casilla de correo 30, 11000, Montevideo, Uruguay}
\author{Jungsang Kim}
\affiliation{Department of Electrical Engineering and the Fitzpatrick Institute for Photonics, Duke University, Durham, North Carolina 27708, USA}
\author{Daniel J. Gauthier}
\affiliation{Department of Physics, The Ohio State University, 191 West Woodruff Ave., Columbus, Ohio 43210 USA}


\date{\today}

\begin{abstract}
	We investigate experimentally a cascade of temperature-compensated unequal-path interferometers that can be used to measure frequency states in a high-dimensional quantum distribution system. In particular, we demonstrate that commercially-available interferometers have sufficient environmental isolation so that they maintain an interference visibility greater than 98.5\% at a wavelength of 1550 nm over extended periods with only moderate passive control of the interferometer temperature ($< \pm0.50~^{\circ}$C). Specifically, we characterize two interferometers that have matched delays: one with a free-spectral range of 2.5~GHz, and the other with 1.25~GHz. We find that the relative path of these interferometers drifts less than 3~nm over a period of one hour during which the temperature fluctuates by $<\pm$0.10~$^{\circ} $C. The error in our measurement is largely dominated by the small drift in the frequency and power of the stabilized laser used to perform the measurement. When we purposely heat the interferometers over a temperature range of 20-50~$^{\circ}$C, we find that the temperature sensitivity is different for each interferometer, likely due to slight manufacturing errors during the temperature compensation procedure. Over this range, we measure a path-length shift of 26 $\pm$ 9~nm/$^{\circ} $C for the 2.5~GHz interferometer. For the 1.25 GHz interferometer, the path-length shift is nonlinear and is locally equal to zero at a temperature of 37.1~$^{\circ}$C and is 50 $\pm$ 17 nm/$^{\circ} $C at 22~$^{\circ}$C. With these devices, we realize a cascade of 1.25~GHz and 2.5~GHz interferometers to measure four-dimensional classical frequency states created by modulating a stable and continuous-wave laser. We observe a visibility $>99\%$ over an hour, which is mainly limited by our ability to precisely generate these states. Overall, our results indicate that these interferometers are well suited for realistic time-frequency quantum distribution protocols. 
\end{abstract}


\maketitle

\section{Introduction}
Quantum distribution (QKD) allows two authenticated parties, Alice and Bob, to share a random key that is secured using the fundamental properties of quantum mechanics \cite{BarnettBook}. The field has progressed rapidly in the last two decades, where most practical QKD protocols encode information in two-dimensional (qubit) states of a photon, such as polarization or relative phase. Today, state-of-the-art QKD systems can generate a finite-length secure key at a rate of megabits per second \cite{Walenta14, Lucamarini13, CharlesDecoy} and at distances over 300 kilometers \cite{Boris2014, Yin2016}, albeit at lower rates.

Despite the significant progress in realistic implementations, the key generation rates in qubit-based protocols are constrained by experimental non-idealities, such as the rate at which the quantum photonic states can be prepared or the deadtime of single-photon-counting detectors. Moreover, in long-distance QKD, a large fraction of the information-carrying photons are lost in the quantum channel due to absorption or scattering. Such physical and practical limitations inspire new QKD protocols that can outperform qubit-based protocols in both secure key rate and distance. 

A class of protocols that is predicted to provide better key rates with higher tolerance against errors involves encoding information in qudit states of photons \cite{Bechmann01, Cerf02, Valero09, Valero10}. We denote the dimension of the Hilbert space describing the quantum states by $d$, where $d=2$ indicates a qubit and $d>2$ indicates qudits. In high-dimensional schemes, information is encoded in various degrees-of-freedom of the photon, such as polarization, time-frequency \cite{Brougham13, Shapiro13, Nunn13, Shapiro14, Dan14, MIT2015, Brougham15, Brecht15}, orbital angular momentum \cite{Boyd14}, or a combination of these \cite{Kwiat05}. High-dimensional protocols have two primary advantages over qubit protocols. First, they allow multiple bits of information to be encoded on a single photon, hence increasing the channel capacity. For some high-dimensional protocols, this can increase the secure key rate for high-loss channels, whereas others can improve the rate when the system is limited by detector saturation. Second, high-dimensional protocols are more robust to channel noise \cite{Bechmann01, Cerf02, Howell07} and can tolerate a higher quantum bit error, thus achieving secure communication at longer distances than qubit protocols \cite{Valero10}. 

We consider a high-dimensional time-bin encoding protocol where information is encoded in frames of time bins and the presence of an eavesdropper is monitored by transmitting mutually unbiased basis (MUB) states with respect to time. For high-dimensional time-bin states, one choice for a MUB is to use states that are the discrete Fourier transform of the temporal states within a frame, known as frequency or phase states \cite{Brougham13}.

The primary challenge of implementing this high-dimensional protocol is measuring the frequency states. One proposed method for measuring frequency states is to use a cascade of $d-1$ unequal-path (time-delay) interferometers \cite{Brougham13}. Experimentally, stabilizing the path difference in the interferometers to sub-wavelength distance scales over long periods of time is challenging due to environmental disturbances, such as temperature, pressure, and vibration, especially for large path differences. While active stabilization of the interferometers is possible, it greatly increases the system complexity.

An alternative approach is to use passively-stabilized interferometers, which have been developed over the last decade by the optical telecommunication industry for use in classical phase- and frequency-domain protocols \cite{Hillerkuss10, Hillerkuss11}. One design principle for addressing the thermal change of the path length is to adopt athermal design, where materials with different thermal expansion coefficients are used to achieve temperature compensation \cite{Thuillier85, Gault85, DLIPatent}. Furthermore, the sensitivity to pressure is reduced by hermetically sealing the interferometer, and the vibration sensitivity is reduced using a compact package. 

Recently, other high-dimensional time-frequency QKD protocols based on continuous~\cite{Shapiro13} or discrete variables~\cite{Nunn13} have been proposed using different approaches for measuring the frequency states based on dispersive optics, for example. The idea of these protocols is to create frequency states using a dispersive media such as a fiber Bragg grating that chirps a single-photon wavepacket, which is decoded by Bob using a conjugate-dispersion fiber Bragg grating followed by single-photon detectors, sometimes combined with a delay interferometer. However, matching the dispersion of the Bragg gratings at the transmitter and receiver in the presence of environmental disturbances is also challenging. 

The primary purpose of this paper is to characterize the stability of commercially-available, passively-stabilized, unequal-path-length interferometers and assess their feasibility for detecting high-dimensional quantum photonic frequency states. By using these athermal interferometers, it is possible to eliminate the need for an active relative phase stabilization of Alice and Bob's interferometers often accomplished by sending strong coherent states between them \cite{Muller96, Ribordy98}, thus eliminating possible Trojan-horse attacks by an eavesdropper \cite{Valero09, Makarov2014}. In addition, these interferometers may find application in coherent one-way and differential-phase-shift QKD protocols \cite{Valero04, Tobias12, Yamamoto02}, or in checking for coherence across many pulses as required for the round-robin protocols \cite{Koashi14, Ma15RR, Koashi15}. Finally, our setup can also be used to show a violation of Bell's inequality in high-dimensional systems \cite{Franson89, Erika11}. 

The paper is organized as follows. In Sec.~\ref{TFProtocol}, we give a brief description of two- and four-dimensional time-frequency QKD protocols and discuss how frequency states can be measured with a cascade of interferometers. In Sec.~\ref{DelayLineInterferometers} we discuss the basic design and stability (Sec.~\ref{InterferometerPerformance}) of these interferometers. In Sec.~\ref{FourDimensionalQKD}, we demonstrate frequency measurement with classical coherent pulses, and we summarize our work and discuss potential future applications in Sec. 6. 

\section{Time-Frequency QKD Protocol}\label{TFProtocol}
\begin{figure}[htb]
	\begin{center}
		\includegraphics[width = 0.35\textwidth]{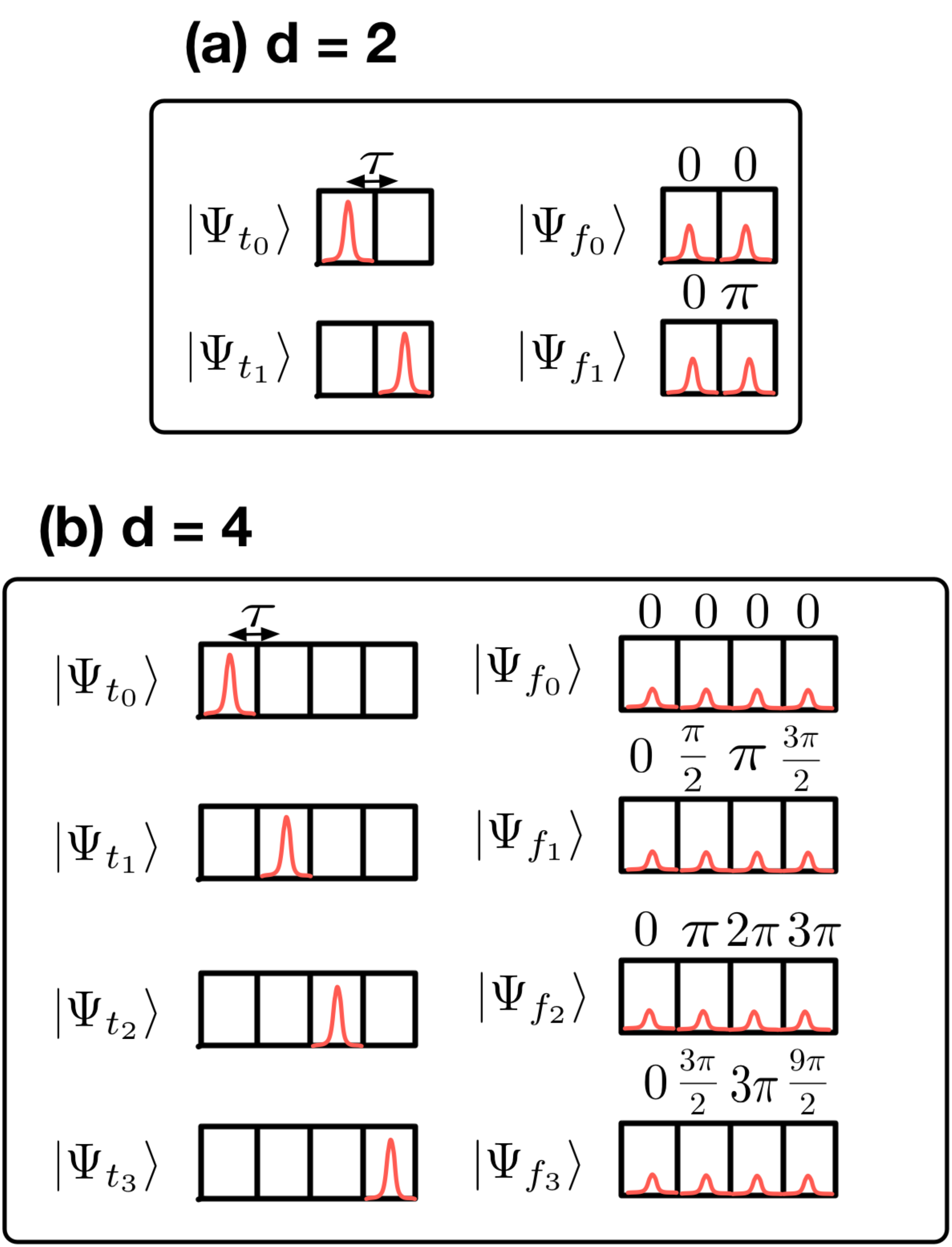}
		\caption{Illustration of temporal (left) and frequency (right) states in terms of the wavepacket temporal positions for (a) $d = 2$ and (b) $d = 4$. The wavepacket peak shapes within each time bin represent the probability density of the photonic wavepacket. The relative phase of the wavepackets for the frequency states is labeled above each time bin.}
		\label{States}
	\end{center}
\end{figure}

We consider the two-basis time-frequency protocol proposed in Ref.~\cite{Brougham13}, which is based on an entangled single photon source, where Alice and Bob share a pair of hyper-entangled photons \cite{Kwiat97}. For simplicity, the discussion below is restricted to the equivalent prepare-and-measure scenario, where Alice prepares and sends single photon states, and Bob measures the incoming states in one of two MUBs. In this protocol, time is discretized into bins of width $\tau$ and grouped into frames of $d$ contiguous time bins. A temporal state $|\Psi_{t_n}\rangle$ is created when the photonic wavepacket is prepared in a single time bin within a frame, which encodes $\mathrm{log}_2 d$ bits. For the frequency states $|\Psi_{f_n}\rangle$, the photonic wavepacket has an equal-height peak in every time bin within the frame and each wavepacket has a distinct relative phase. Here, the integrated probability over a frame is held constant for all of the time and frequency states. Figure~\ref{States} illustrates the $d=2$ and $d=4$ states.

In greater detail, the temporal states can be written as $|\Psi_{t_n}\rangle = a_n^{\dagger}|0\rangle$, where $a_n^{\dagger}$ is the field creation operator acting on a vacuum state in the in $n^{th}$ temporal mode. Consequently, the frequency states can be written as \cite{Cerf02}
\begin{eqnarray}
|\Psi_{f_n}\rangle = \frac{1}{\sqrt{d}} \sum\limits_{m= 0}^{d-1} \exp\left(\frac{2\pi i n m}{d}\right)|\Psi_{t_m}\rangle, ~~~n = 0, ... d-1 
\end{eqnarray}
%
%
which is a natural extension of the BB84 temporal qubit states to higher dimension. 

\begin{figure*}[htp]
		\includegraphics[width = 0.8\textwidth]{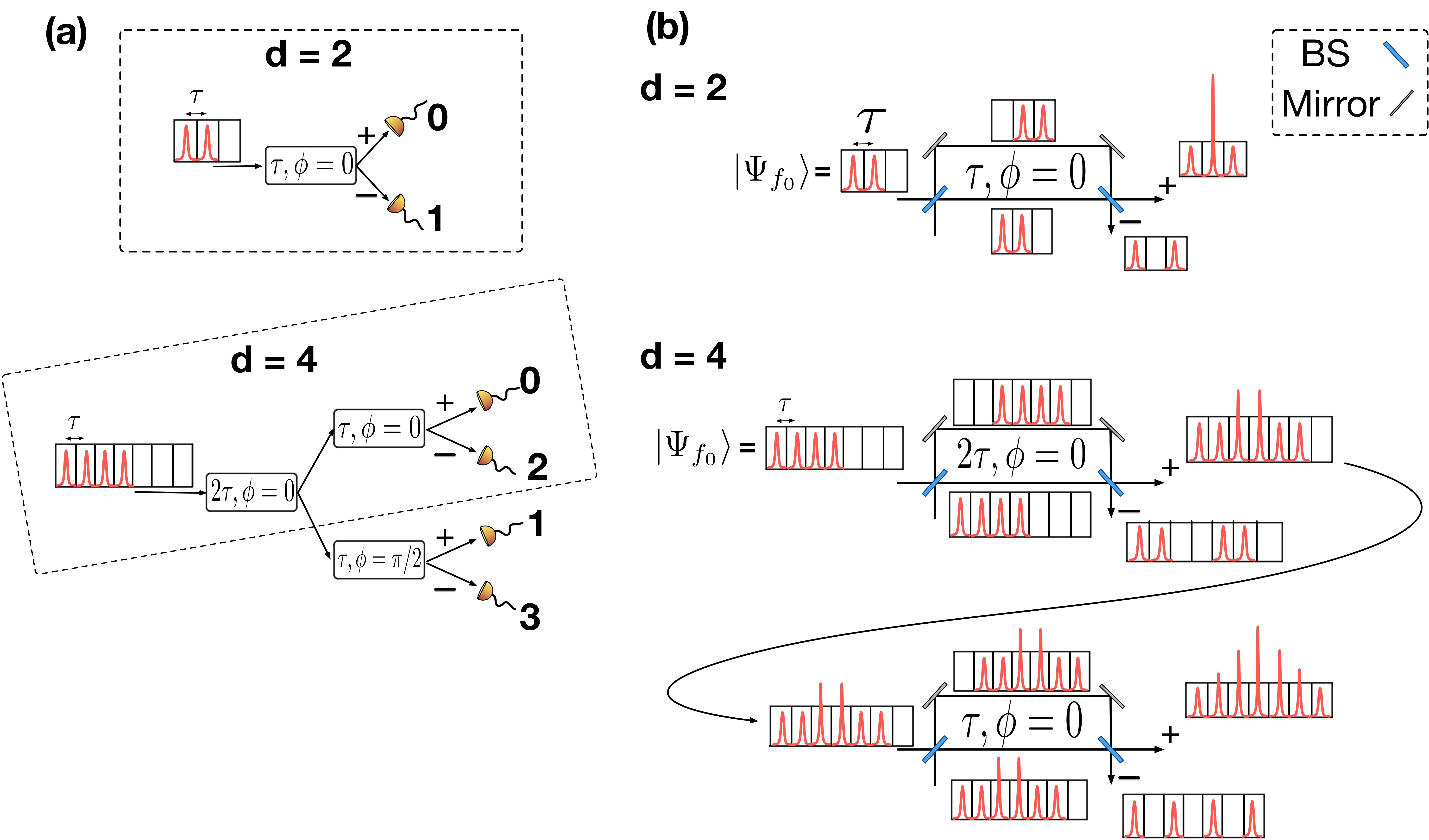}
		\caption{Schematic illustration of the frequency-state measurement technique for $d = 2$ (top panels) and $d = 4$ (bottom panels). (b) Frequency state waveforms at different locations in the interferometers.}
		\label{FrequencyBasisMeasurement}
\end{figure*}

In a typical experimental implementation of a discrete-variable time-frequency QKD system, an attenuated laser is used to generate the photonic wavepackets with a mean photon number of the order of 1, and a generalized decoy state protocol is used to put tight bounds on the fraction of wavepackets that have more than one photon \cite{Lodecoy05, CharlesDecoy}. The temporal states can be generated by on/off encoding of a continuous-wave laser with a high-contrast intensity modulator. They can be measured directly with a single-photon detector with a jitter much less than $\tau$ and the event recorded with a high-resolution time tagger. The frequency states can be generated using a combination of phase and intensity modulators or with a cascade of delay interferometers considered here. 

One scheme for measuring the frequency states is shown in Fig.~\ref{FrequencyBasisMeasurement}a for the case $d=2$, where we assume that the wavepacket peaks have a width much less than $\tau$. Here, the relative phase difference is 0 for $|\Psi_{f_0}\rangle$ or $\pi$ for $|\Psi_{f_1}\rangle$. As will become apparent below, an empty `guard' prevents overlap of wavepacket peaks from neighboring frames when operating the QKD system at a high rate. A detailed analysis (not presented here) shows that this overlap does not affect the quantum bit error rate substantially and hence the guard bins are not necessarily needed. For clarity, we include the use of the guard bins in the discussion below.

In a time-delay interferometer, an incoming beam is split equally by a 50-50 beamsplitter and directed along two different paths and recombined at a second 50-50 beamsplitter where the wavepackets interfere. The difference in path between the two arms of the interferometer is denoted by $\Delta L = \Delta L_0 + \delta L$, where $\Delta L_0$ is the nominal path difference. Here, $\delta L \ll \Delta L_0$ is a small path difference that allows us to make a fine adjustment to the transmission resonances of the interferometer and is proportional to the phase $\phi = k\delta L $, where $k$ is the magnitude of the wavevector of the wavepacket.

For $d=2$, only a single time-delay interferometer is required with $\Delta L = c \tau$, corresponding to a free-spectral range (FSR) $c/\Delta L$, where $c$ is the speed of light and $\phi$ is set to zero. When the state $|\Psi_{f_0}\rangle$ is incident on the interferometer, the wavepacket traveling along the long path is delayed by $\tau$ with respect to the wavepacket traveling along the short arm. After the second beamsplitter, the wavepacket originally occupying two time bins now occupy three (and hence explains the need for a `guard' bin), where only the wavepacket peak at the center of each frame interferes constructively (destructively) for the $+$ ($-$) port. The earliest and the latest wavepacket peaks of the state do not interfere at the second beamsplitter and hence do not directly give information about the frequency state. The situation is reversed when the state $|\Psi_{f_1}\rangle$ is incident on the interferometer (not shown). 

For $d = 4$, one possible approach for measuring the frequency states uses a cascade of three time-delay interferometers as shown in the lower panel to Fig.~\ref{FrequencyBasisMeasurement}a. The first interferometer has a path difference of $2 c \tau$, while the two interferometers connected to the output ports of the first interferometer have a path difference of $c \tau$. The phase of the interferometer connected to the $+$ ($-$) port of the first interferometer is set to $\phi = 0~(\pi/2)$ whose outputs allow us to measure the frequency states $|\Psi_{f_0}\rangle$ and $|\Psi_{f_2}\rangle$ ($|\Psi_{f_1}\rangle$ and $|\Psi_{f_3}\rangle$). 

The frequency states for $d=4$ have four contiguous time bins occupied by wavepacket peaks of different relative phases and require three guard bins. When the state $|\Psi_{f_0}\rangle$ is incident on the first interferometer, as illustrated in Fig.~\ref{FrequencyBasisMeasurement}b, the wavepackets are shifted temporally by $2\tau$ when they arrive at the second beamsplitter and there is constructive (destructive) interference for the wavepackets in the two middle time bins at the $+$ ($-$) port of the interferometer. The two outer wavepackets do not experience interference. These two sequences of wavepackets are directed to the second set of interferometers of the next layer in the cascade. For simplicity, we only describe the interferences that take place in the interferometer with $\phi = 0$, indicated by the dashed box shown in the lower panel of Fig. ~\ref{FrequencyBasisMeasurement}a.

At the $+$ ($-$) output port of the second interferometer (lower panel of Fig.~\ref{FrequencyBasisMeasurement}b), all 7 time bins are occupied with the highest (lowest) probability for photon occupation for the time bin in the middle of the frame, which is due to constructive (destructive) interference of all four wavepacket peaks of the incident state. The other occupied time bins give information about interference of a subset of the incident wavepacket peaks except for the outermost time bins where no interference occurs. Thus, it is advantageous to measure all of the central 5 time bins because they each measure different aspects of the coherence of the incident wavepacket peaks. Because of the possibility of measuring the coherence among different sets of wavepacket peaks, the cascade of interferometers might also find use in the recently developed round-robin QKD protocol \cite{Koashi15}.

A similar analysis shows that the central time bin of each of the four outputs from the interferometer cascade are directly related to each of the frequency states. That is, constructive interference occurs in output port $n$ when the state $|\Psi_{f_n}\rangle$ is incident on the cascade and destructive interference is observed in the other three ports. The procedure for arbitrary $d$ is given in Ref.~\cite{Brougham13} for which $2d-1$ guard bins are required.

In a time-frequency QKD protocol, the contrast (visibility) of the interference provides an estimate of the error introduced by an adversary (Eve) in the quantum channel. The visibility for a frequency state $|\Psi_{f_n}\rangle$ is defined as
\begin{equation}
\mathcal{V} = \frac{\mathcal{P}_+ - \mathcal{P}_-}{{\mathcal{P}}_+ + \mathcal{P}_-}, \label{Vis}
\end{equation}
where $\mathcal{P}_+$ is the probability of detecting the photon in the expected bright port $n$ in the central time bin and $\mathcal{P}_-$ is the probability of finding the photon in any of the other ports in the same central time bin. The interference visibility is limited in a real device by the accuracy of the beamsplitting ratio, differential loss in the paths, and larger beam diffraction in one path in comparison to the other. Eve's interaction with the quantum states result in a loss of temporal coherence of the frequency states and thus results in a reduction of $\mathcal{V}$. A security analysis determines the maximum error rate that can be tolerated and hence the minimum value of $\mathcal{V}$. 

\section{Time-Delay Interferometers}\label{DelayLineInterferometers}
\begin{figure}[htp]
	\begin{center}
		\includegraphics[width = 0.35\textwidth]{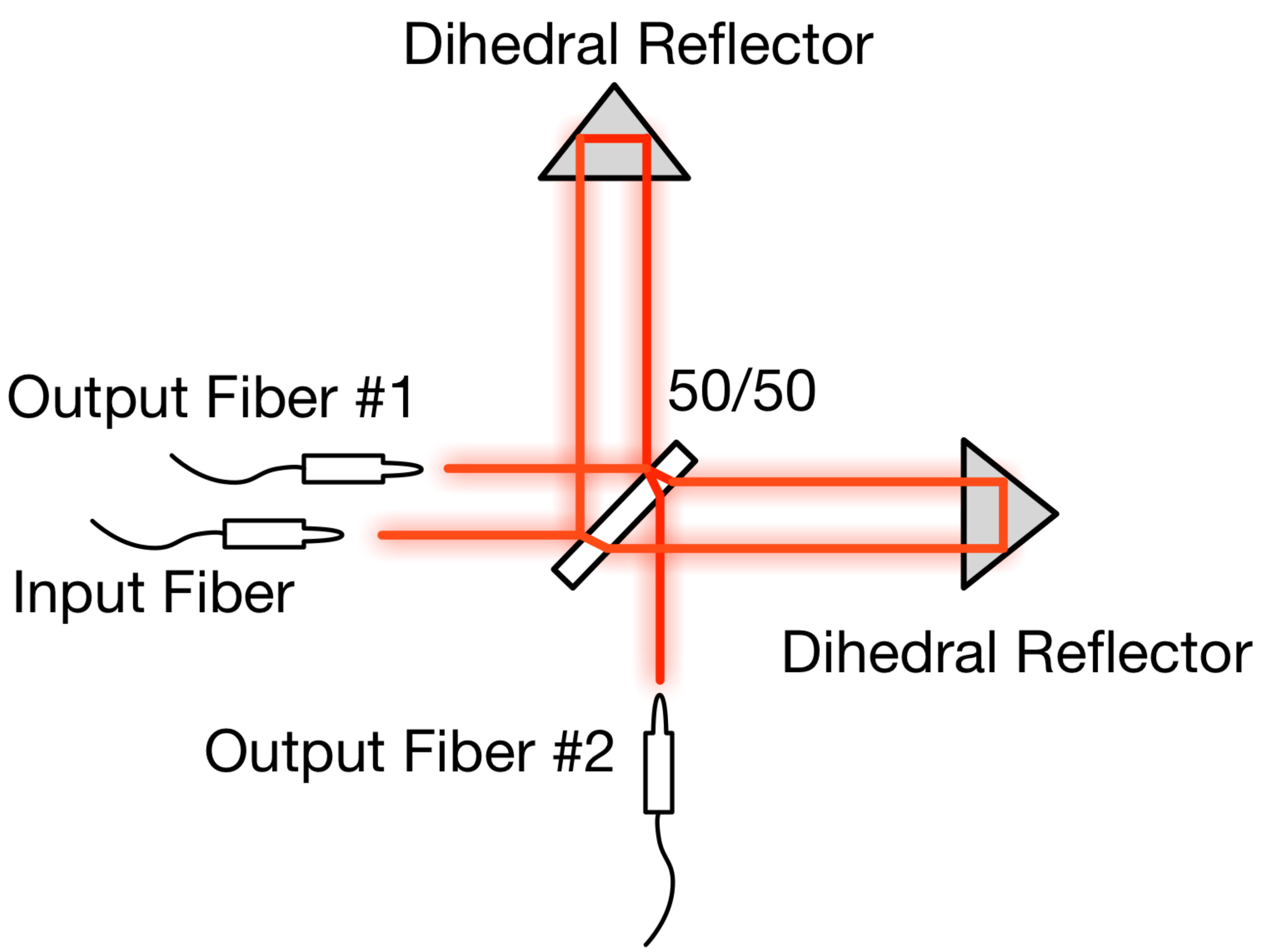}
		\caption{Illustration of the internal components of a typical delay-line interferometer \cite{Ludvoic}.}
		\label{Kylia_Schematic}
	\end{center} 
\end{figure}

The delay line interferometers used in the experiment are manufactured by Kylia and are of the Mach-Zehnder type, but use a folded design reminiscent of a Michelson interferometer with displaced input and output beams \cite{Thuillier85, Gault85} to make the design more compact and simpler (Fig.~\ref{Kylia_Schematic}). Here, the incoming beam of light is split into two unequal paths using a 50-50 beamsplitter, displaced by the dihedral reflectors, recombined at the same beamsplitter, and are directed to two output ports. The overall phase $\phi$ of the interferometer is adjusted by changing the optical path of one arm of the interferometer relative to the other using a resistive heater placed near one of the reflectors. The path change is proportional to the power delivered to the resistor; applying $\leq$3~V results in a path length change of approximately one FSR. Our devices are designed to operate over the classical optical telecommunication C-band and we evaluate their performance at 1550 nm near the middle of the band.

The stability of these interferometers against environmental changes depends on the thermal compensation method. While the Kylia design is proprietary, typical temperature-compensated delay-line interferometers use materials with a low coefficient of thermal expansion and an optimized selection of glasses and air paths for thermal and chromatic compensation \cite{Thuillier85, Gault85,DLIPatent}. The Kylia devices are realized using ultra-low-expansion optical glass components and base plate and packaged inside a hermetically sealed aluminum housing, which stabilizes them against environmental temperature and pressure, respectively. 

\section{Interferometer Performance} \label{InterferometerPerformance}
The change in path of the interferometer $\delta L$ as a function of temperature $T$ is typically specified by a temperature-dependent path-length shift (TDPS), but the time scale over which this characteristic is measured is usually not specified and the use of only a single metric assumes that it is independent of $T$. As we show here, such a simple metric is not sufficient to adequately describe the relation between $\delta L$ and the change in temperature $\Delta T$. This is due to the fact that the outer package is made from aluminum, the internal interferometer is constructed from potentially different types of glass, and the input and output fibers must pass through the outer aluminum package. These materials have vastly different thermal conductivities and heat capacities, and the detailed design of the thermal link between them is proprietary. As described below, we observe two different time scales for the TDPS and that it can be a nonlinear function of $T$, indicating that the standard commercial specification is insufficient.

\begin{figure}[htp]
	\begin{center}
		\includegraphics[width = 0.45\textwidth]{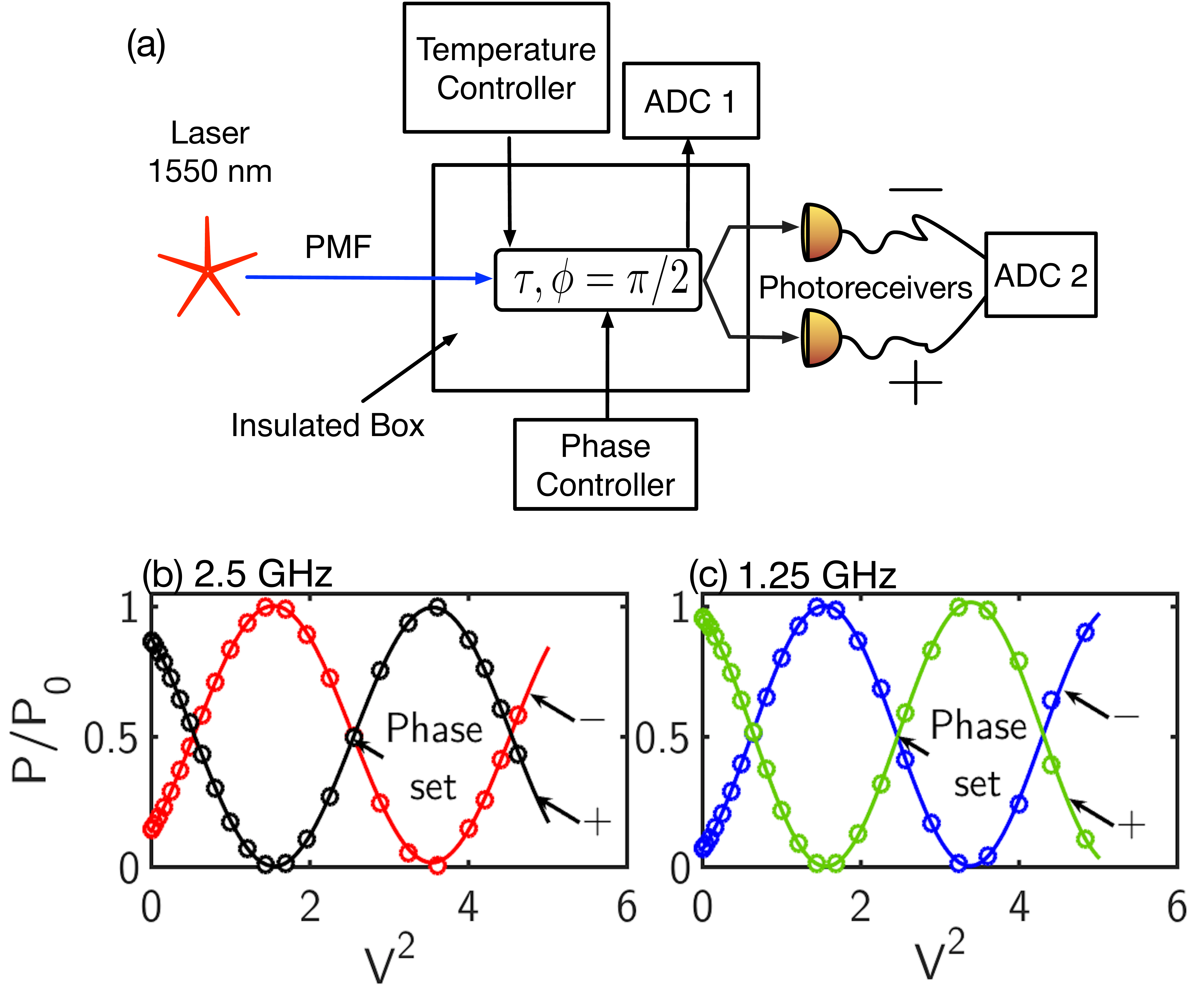}
		\caption{(a) Schematic of the setup used to evaluate the performance of the interferometers. A 200 $\mu$W continuous wave laser beam is injected into the interferometer using a polarization maintaining fiber (PMF). The temperature of the interferometer is monitored using multiple thermocouples placed at different locations on the interferometers and digitized using ADC 1 (National Instruments NI 9239). The output powers are recorded using two photoreceivers (New Focus 2011) and digitized using ADC 2 (National Instruments NI 9239). The power at the two outputs of the (b) 2.5 GHz and (c) 1.25 GHz interferometers as a function of the square of the voltage applied to the resistive heater.}
		\label{Stability_Experiment}
	\end{center} 
\end{figure}

We investigate the performance of two Kylia delay line interferometers, one with an FSR of 1.25 GHz ($\Delta L_0$ = 24~cm, $\tau$=800~ps) and the other with 2.5 GHz ( $\Delta L_0$ = 12~cm, $\tau$=400~ps). The performance of these devices is characterized by observing the variation in the power of light emerging from one of the output ports when a continuous wave, single-frequency laser beam is injected into the interferometer, as shown in Fig.~\ref{Stability_Experiment}a, for three situations: 1) long-term ($\sim$ an hour) stability in a controlled laboratory environment (temperature control of $\pm$0.1~$^\circ$C); 2) long-term ($\sim$ an hour) visibility in the same laboratory environment; and 3) the TDPS as we vary $T$ between 20 and 50 $^\circ$C. 

Based on the specification of the interferometers (TDPS $<$ 50\% of the FSR over a 0 - 70 $^\circ$C temperature range), we expect the shift in resonance frequency of the interferometers to be less than 10 MHz for $T< 0.5$ $^\circ$C, which is typical in a laboratory environment. In order to measure such a small variation, we use a frequency-stabilized laser (Wavelength Reference Clarity-NLL-1550-HP locked to an HCN line and operating in the `Line Narrowing' mode) with an absolute accuracy of $\leq~\pm0.3$~pm and a specified long-term root-mean-square (RMS) frequency stability better than 1~MHz.

For all measurements, the interferometers are placed in a thermally insulated box and allowed to equilibrate for $\sim$2 hours with a mean initial temperature of 21.3~$\pm$~0.3~$^{\circ}$C. For the stability and TDPS measurements at a nominally constant $T$, the phase of the interferometer is set at the beginning of the equilibration process to the steepest slope of the interference fringe ($\phi = \pi/2$), as shown in Fig.~\ref{Stability_Experiment}b. If $\phi \neq \pi/2$ at the end of the equilibration process, a small change is made to bring it back to this point. For the visibility measurement, $\phi$ was set to zero to place it at an interference maximum. For the TDPS measurements over a wider range, $T$ is set to between 20 and 50~$^{\circ}$C using heating tapes wrapped around the device, which are connected to a variable voltage supply. 

\subsection{Stability at nominally constant temperature}
The optical power emerging from the $\pm$ ports of an ideal (high-visibility) time-delay interferometer is given by
\begin{equation}\label{Power}
P_{out,\pm} = \frac{\alpha P_{0}}{2}(1\pm\cos(k\Delta L)),
\end{equation}
where $P_{0}$ is the power at the input of the interferometer and the parameter $\alpha \in \{0,1\}$ represents the reduced transmission due to insertion loss of the interferometer. We find that the predominant contribution to the variation in $P_{out,\pm}$ arises from imperfect thermal compensation and hermetic sealing of the device, giving rise to a change in $\delta L$. The power at the output is given by
\begin{eqnarray}\label{Drift}
\frac{2P_{out,\pm}}{\alpha P_{0}} = (1\pm\cos(k(\Delta L_0 + \delta L))) \nonumber \\
= (1\pm\cos(\phi \pm k \delta L)), \label{1}\\
= (1\mp\sin(k\delta L)) \label{2}.
\end{eqnarray}
where we insert the phase $\phi \equiv k\Delta L_0 = \pi/2$ between Eq.~\ref{1} and Eq.~\ref{2}. Equation~\ref{2} relates the output power of the interferometer initially set at $\phi=\pi/2$ to $\delta L$ assuming a stable laser frequency and we use it to estimate the drift of the interferometer. 

The emitted power can also change due to other physical effects, which will cause us to incorrectly associate a change $\delta L$ with a change in $P_{out,\pm}$. We use various methods to account for this systematic error in our measurement. To account for variation in the incident laser power $P_0$ (typically below 0.01\%), we place a 50-50 fiber beamsplitter just before the interferometer with one output defining the reference laser power $P_{r}(t)$, while the other output is directed to the interferometer. The ratio between the peak power emitted by one output of the interferometer and the reference power is given by $\alpha$. 

\begin{figure}[htp]
	\centering
	\includegraphics[width = 0.45\textwidth]{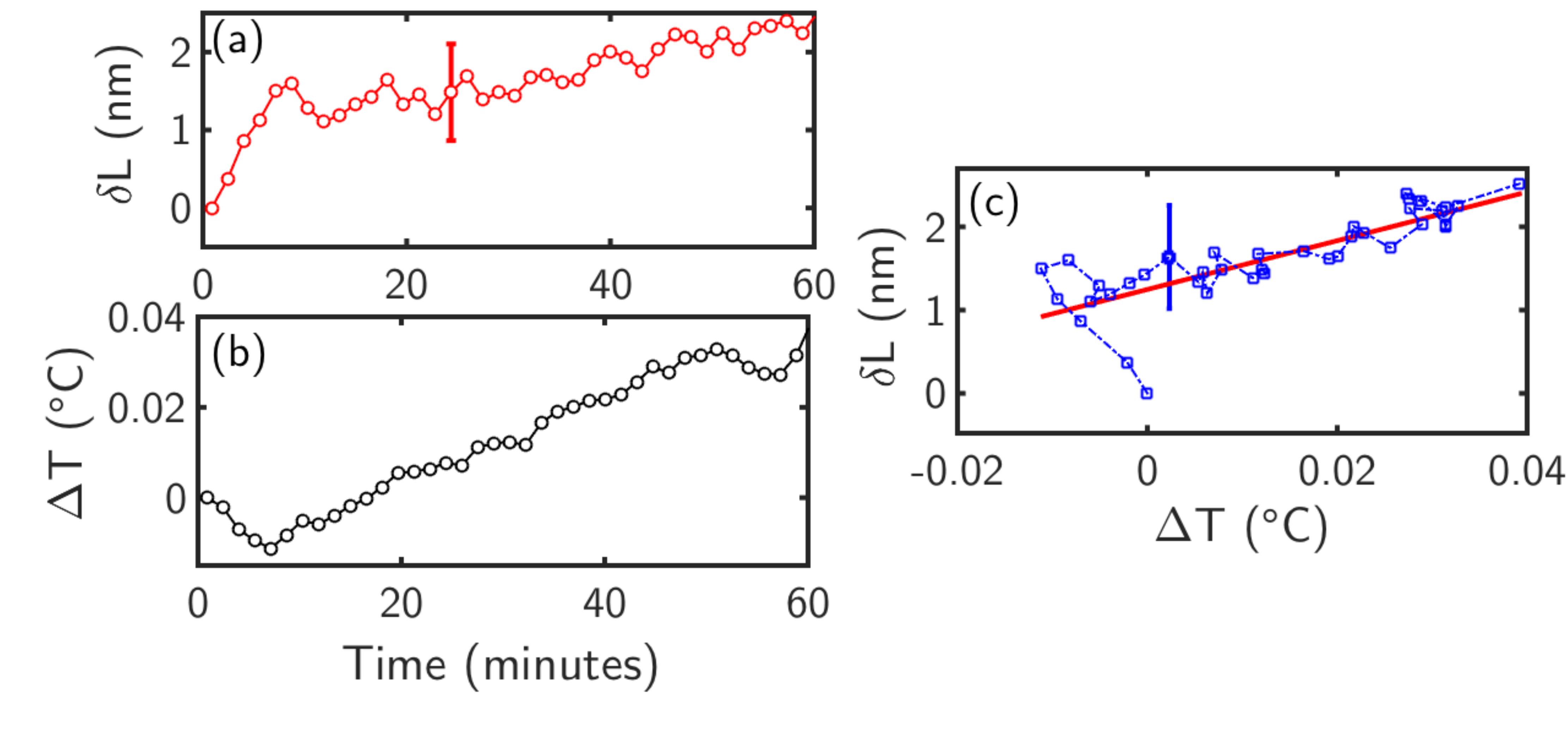}
	\caption{(a) The path length drift of the 2.5 GHz interferometer measured over an hour. (b) The corresponding temperature variation monitored over the same period of time. (c) The path length drift from (a) plotted as a function of the temperature variation from (b).}
	\label{Drift25GHz}
\end{figure}

For the stability measurements, we find that both the 1.25~GHz and 2.5~GHz interferometers display an apparent drift of less than 3~nm over an hour if the temperature of the environment is stabilized to $\pm0.1~^\circ$C as it is in the cardboard enclosure. Figure~\ref{Drift25GHz}(a) shows one such measurement of $\delta L$ (extracted from the data using Eq.~\ref{2}) for the 2.5~GHz interferometer, with the corresponding change in temperature in Fig.~\ref{Drift25GHz}(b). We observe that $\delta L$ is not fully correlated with $\Delta T$ (0.8 correlation coefficient). The lack of stronger correlation can be attributed to two additional factors that affect the apparent drift of the interferometer. 

First, there is a contribution to the drift of the interferometer due to laser frequency fluctuation. We expect the drift of the frequency-stabilized laser to be better than 1~MHz over an hour, which corresponds to an apparent path-length change of 0.62 nm for the 2.5~GHz interferometer as indicated by the error bar in Fig.~\ref{Drift25GHz}(a) and Fig.~\ref{Drift25GHz}(c). To estimate the contribution of laser drift to this data set, we fit it to a linear function as indicated by the red line. We attribute the finite slope of the line as arising from the change in path of the interferometer, which is $1.2\pm 0.1$~nm. This is likely an upper bound to the actual path-length change given that it is within the range of the specified laser-frequency drift. 

From this data, we also determine the root-mean-squared error (RMSE) between the data and the fit. The measured RMSE of 0.32 nm corresponds to a possible laser frequency variation of 0.51~MHz, well within the specified deviation of $<$ 1~MHz and thus we attribute these smaller-scale fluctuations in the data to the laser-frequency drift. Clearly, it is evident that the laser frequency variation is a significant contribution to the apparent path-length change of the interferometer.

A second factor that can give rise to an imperfect correlation between $\delta L$ and $\Delta T$ is the fact that we measure $T$ of the aluminum outer package, which may not reflect the actual temperature of the substrate and optics housed inside the aluminum package (see discussion below). The effect of such a lag on this data is difficult to determine from measurements over a such a small temperature change. To address this issue, we conduct a set of measurements for a larger temperature range as discussed in Sec.~\ref{Sec:TDPS} below. 

Figure~\ref{Drift125GHz} shows similar plots for the 1.25~GHz interferometer. For this particular run, the temperature change of the device ($\sim 0.01~^\circ$C) is much less than when we collected data for the other interferometer. In Fig.~\ref{Drift125GHz}(a), we observe that the interferometer apparently drifts substantially in the first $\sim$ 20~minutes of the run, and then stabilizes to within $\sim$ 1.2~nm thereafter. Again, there is little correlation between $\delta L$ and $\Delta T$ (-0.03 correlation coefficient). Following a similar procedure described above, we find that a straight line fit to the data shown in Fig.~\ref{Drift125GHz}(c) has a slope of zero, implying no path length change over this temperature range. Furthermore, the RMSE between the linear fit and the data corresponds to a path difference of 0.34~nm and can be attributed to a 0.27~MHz drift in the laser frequency, well within the specification of the laser. We note that the 1~MHz specification of the laser-frequency now translates to a 1.24~nm change in path-length as indicated by the error bar in Fig.~\ref{Drift125GHz}(b) and Fig.~\ref{Drift125GHz}(b).

\begin{figure}[htp]
	\centering
	\includegraphics[width = 0.45\textwidth]{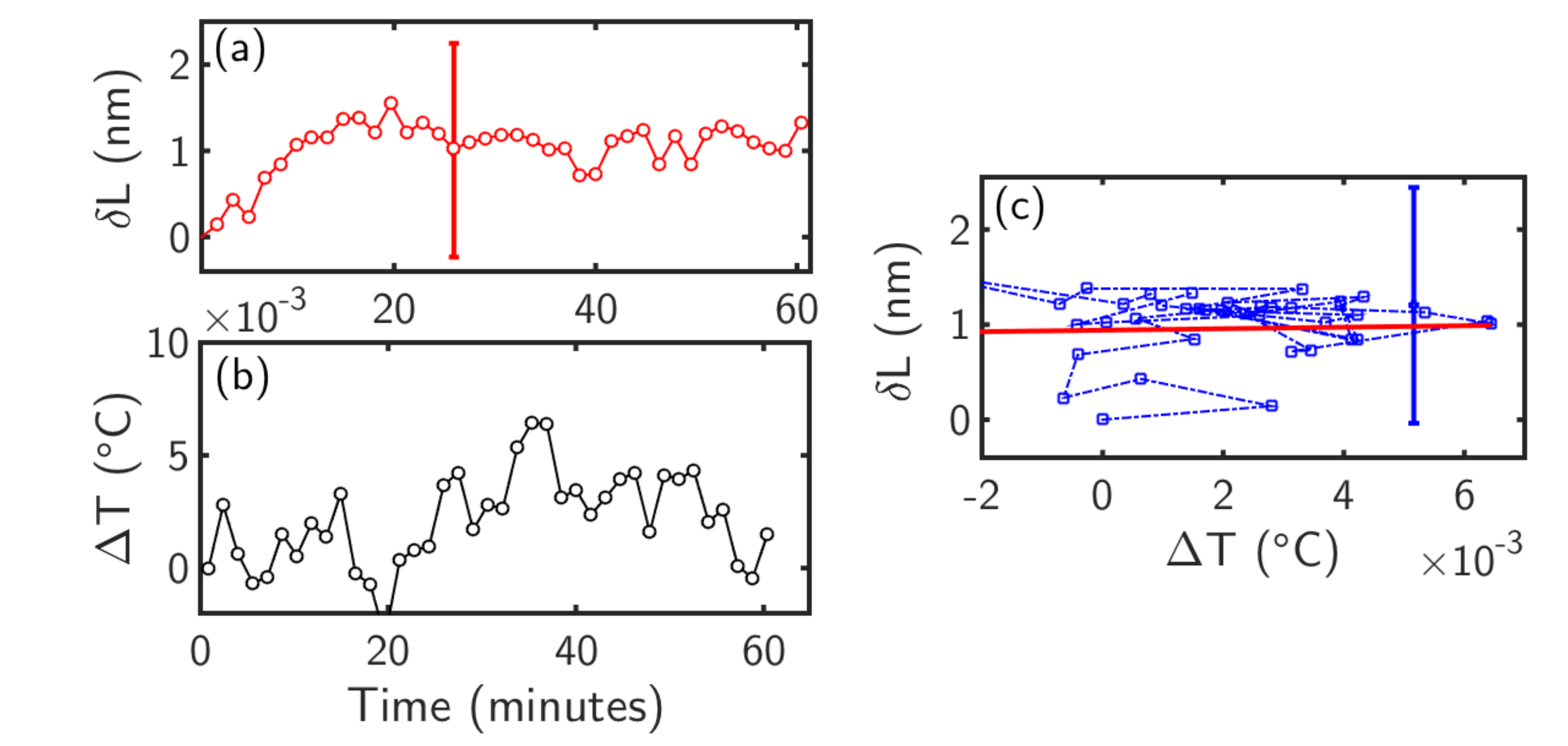} 
	\caption{(a) The path length drift of the 1.25 GHz interferometer for three independent runs measured over an hour. (b) The corresponding temperature variation monitored over the same period of time. (c) The path length drift from (a) plotted as a function of the temperature variation from (b).}
	\label{Drift125GHz}
\end{figure}

We performed a similar analysis on several independent data sets collected for both the interferometers and observed similar stability. Specifically, we observe that $\delta L <$ 3~nm over an hour if the interferometers are stabilized to $<0.1~^\circ$C. This measurement is an upper bound to the true path-length change because the variation of the frequency of the stabilized laser gives rise to a comparable apparent shift. 

\subsection{Visibility}
As discussed in the text related to Eq.~\ref{Vis} above, the interferometer visibility $\mathcal{V}$ is a critical QKD system parameter used to determine an upper bound on the effects due to an eavesdropper. Thus, to extract the largest possible key, it is important to characterize the base-line change in $\mathcal{V}$ due to environmental conditions, which will set a lower limit on the error due to Eve that can be detected. To this end, we inject a continuous-wave, frequency-stabilized laser beam into the interferometer (see Fig.~\ref{Stability_Experiment}a), set $\phi=0$, and monitor the power coming out of the output ports, denoted by $P_{max}$ at the + output port and $P_{min}$ at the - output port. We then determine $\mathcal{V}$ using Eq.~\ref{Vis}, where the probabilities $\mathcal{P_{+}}$ and $\mathcal{P_{-}}$ are replaced with powers $P_{max}$ and $P_{min}$. We do not monitor the laser power in these measurements because the typical variations in laser power ($<$0.01\%) has less than a 0.004\% effect on $\mathcal{V}$. 

\begin{figure}[htp]	
	\centering
	\includegraphics[width = 0.45\textwidth]{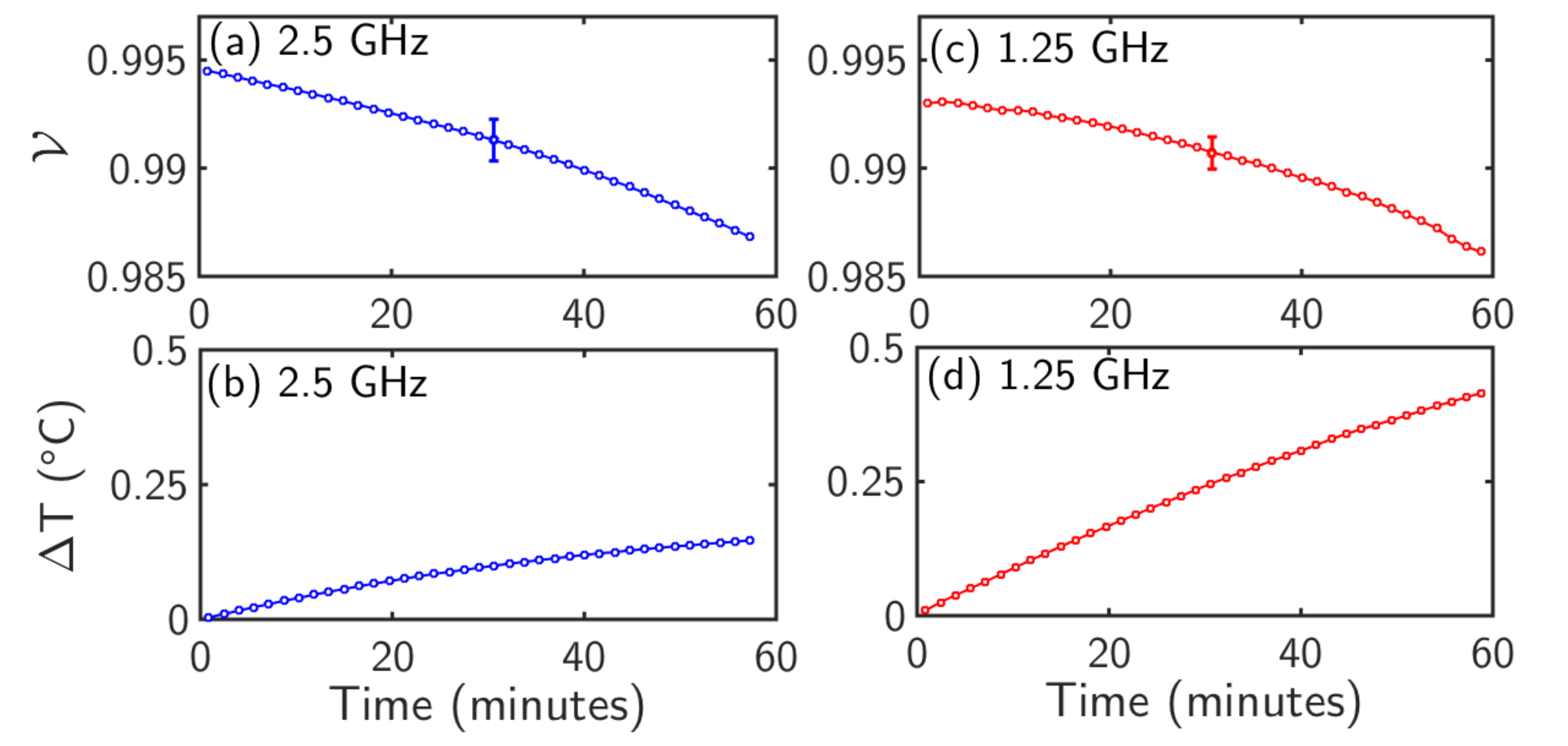} 
	\caption{(a) and (c) The visibility of the 2.5 GHz and 1.25 GHz interferometers measured over an hour, respectively. (b) and (d) The temperature variation of the 2.5 GHz and 1.25 GHz interferometers measured over an hour, respectively.}
	\label{Visibility}
\end{figure}

Figure~\ref{Visibility} shows the temporal behavior of $\mathcal{V}$ and $\Delta T$ for both interferometers measured independently at two different times, each over the course of an hour. We note that the temperature changes are slightly larger ($<\pm$ 0.5 $^\circ$C) than for the stability measurement discussed in the previous section. For both interferometers, we find that the visibilities stay well over 98.5~\% during the entire hour. The error bars indicate the expected change in $\mathcal{V}$ for a typical drift in the laser frequency of 1~MHz. This error was determined by propagating uncertainties and the covariance of the dependent variables $P_{max}$ and $P_{min}$. 

One potentially useful application of the delay interferometers is to perform frequency-state measurements for a wavelength-division multiplexed time-frequency QKD system. To assess the Kylia devices for this application, we use a widely tunable laser (Agilent HP81862A) to measure $\mathcal{V}$ for both interferometers between 1525~nm to 1565~nm (approximately over the entire C-band) in $\sim$5~nm steps. We find that $\mathcal{V} > 99~\%$ across this range. This is consistent with Kylia's specification of $>$ 98.4\% over the wavelength range of 1520-1570~nm. Thus, in a wavelength multiplexed system, a single set of interferometers can be used for the frequency measurement for each wavelength channel and a wavelength demultiplexer can be placed after the interferometers to send each channel to their respective detectors. Therefore, a high data throughput time-frequency QKD system can be realized without using a separate delay-interferometer cascade for each spectral channel. 

\subsection{Wide-Range Temperature-Dependent Path-Length Shift}\label{Sec:TDPS}
To obtain a better estimate of the TDPS that is not as sensitive to the laser frequency fluctuations, we measure $\delta L$ over a wider temperature range, where the path-length shift is expected to be larger. The setup is identical to that described in Sec.~\ref{InterferometerPerformance} above except that we purposefully vary the device temperature in large steps. We collect data at each temperature step for at least 6 hours. After this interval, we heat the device again to a new temperature, repeating this procedure until the total temperature change is $\sim$30~$^{\circ}$C from an initial temperature of $\sim$22$^{\circ}$C. 

\begin{figure}[h]
	\centering
	\includegraphics[width = 0.45\textwidth]{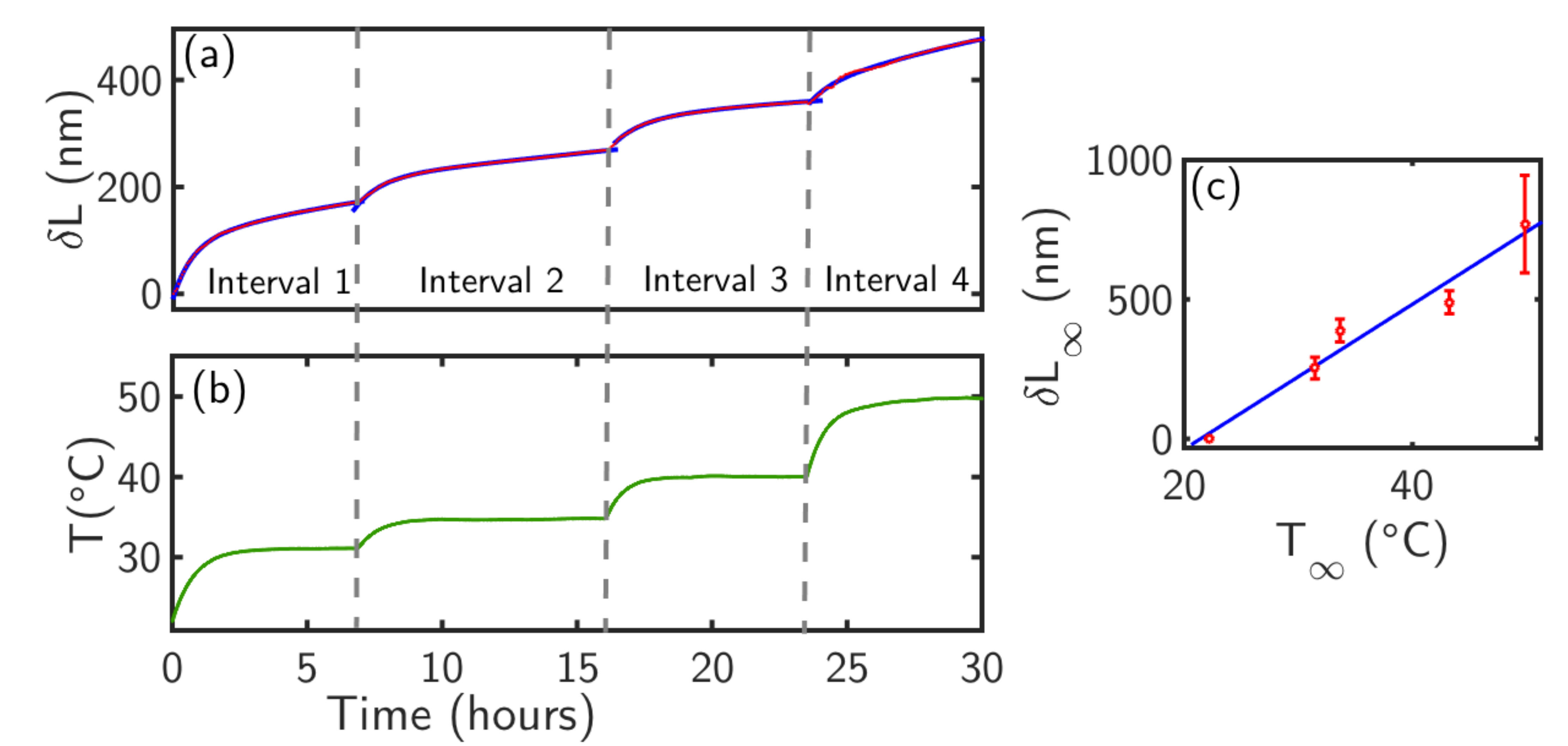}
	\caption{Temperature-dependent path-length shift for the 2.5 GHz interferometer. (a) Variation in the path change as the interferometer is heated in four intervals. The rate constants for the double-exponential fit for intervals 1-4 are (1.4 $\pm$ 0.1, 0.10 $\pm$ 0.04 ), (1.3 $\pm$ 0.1, 0.10 $\pm$ 0.02), (1.3 $\pm$ 0.1, 0.23 $\pm$ 0.04) and (1.6$\pm$0.6, 0.07 $\pm$ 0.06)~hr$^{-1}$, respectively, and the long-time extrapolated path change $\delta L_{\infty}$ are 253 $\pm$ 80, 135 $\pm$ 3, 101 $\pm$ 19, and 280 $\pm$ 170~nm, respectively. (b) Variation in the temperature as the interferometer is heated. The rate constants for the exponential fit for intervals 1-4 are 1.223 $\pm$ 0.007, 1.370 $\pm$ 0.007, 1.48 $\pm$ 0.02, and 1.047 $\pm$ 0.005~hr$^{-1}$, respectively and the extrapolated temperature is 32.35 $\pm$ 0.03, 36.1 $\pm$ 0.2, 41.6 $\pm$ 0.1, and 50.78 $\pm$ 0.03~$^{\circ}$C, respectively. (c) Temperature dependence of the long-time extrapolated path change along with a fit to a straight line.}
	\label{TDFS1}
\end{figure}

Figure~\ref{TDFS1} shows the variation in $\delta L$ with $T$ for the 2.5~GHz interferometer over four heating intervals (intervals indicated by vertical dashed lines). At the beginning of each interval, we observe that the temperature of the aluminum housing increases and then levels off. We find that the data is well described by a single exponential function with a rate constant of 1.28$\pm$0.01 hr$^{-1}$ averaged over the four time intervals (Fig.~\ref{TDFS1}(b)). The data and the fit are overlaid in the figure and are indistinguishable (reduced $\chi^2$= 1.34). From Fig.~\ref{TDFS1}(a), we see that there is a correspondingly rapid increase in $\delta L$, followed by a slower continued rise. We find that the data is fit well by a double exponential with two different rate constants; the fit function is again overlaid with the measurements and are nearly indistinguishable (reduced $\chi^2$= 1.42). The two average rate constants are 1.4$\pm$0.2 hr$^{-1}$ and 0.13$\pm$0.02 hr$^{-1}$. The larger rate constant is similar to that for the rise in $T$, and we attribute this change in path due to coupling between the aluminum housing and the interferometer, likely due to mechanical coupling between the two. The lower heat conductivity of the interferometer glass likely contributes to making the other rate constant so long. 

To estimate the total change in $\delta L$ for each interval even though we do not collect data long enough to reach equilibrium, we use our double-exponential fit to find the long-time limit for the path change, which we denote by $\delta L_{\infty}$, and the single-exponential fit to find T$_\infty$. We note that this is only an estimate because it assumes a change in $\delta L$ that is linear with temperature after a sufficiently long settling time. Figure~\ref{TDFS1}(c) shows $\delta L_{\infty}$ as a function of $T_{\infty}$, which we fit with a straight line. From this fit, we find that the TDPS is $26 \pm 9$ nm/$^{\circ}$C. The TDPS specified by Kylia is 11 nm/$^{\circ}$C, which is clearly smaller than what we estimate for our device. This could be due to imperfect temperature compensation for this device. 

\begin{figure}[h]
	\centering
	\includegraphics[width = 0.45\textwidth]{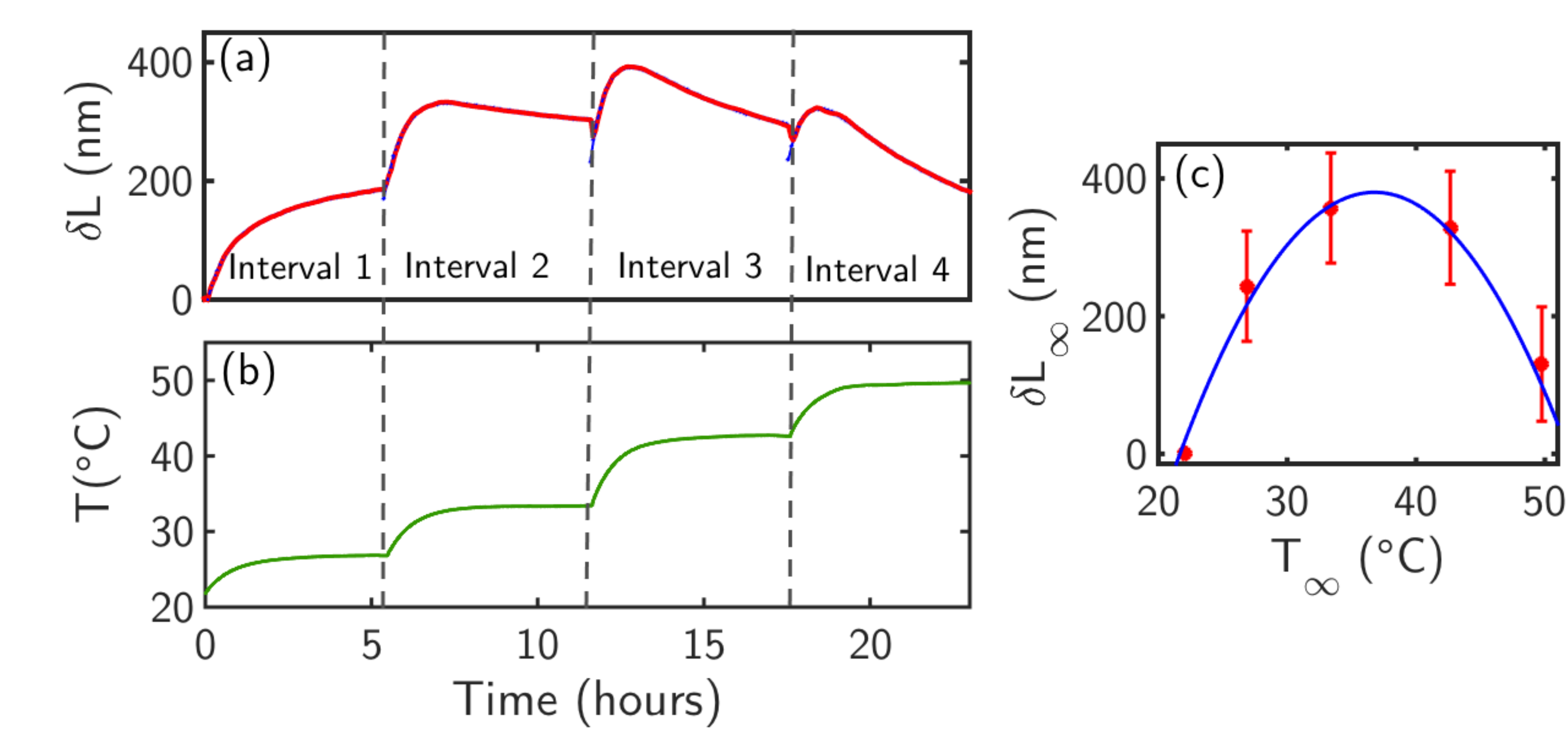}
	\caption{Temperature-dependent path-length shift for the 1.25 GHz interferometer. (a) Variation in the path change as the interferometer is heated in four intervals. The rate constants for the double-exponential fit for intervals 1-4 are (1.4 $\pm$ 0.2, 0.2 $\pm$ 0.1), (1.2 $\pm$ 0.2, 0.6 $\pm$ 0.2), (1.9 $\pm$ 0.1, 0.28 $\pm$ 0.04) and (1.8 $\pm$ 0.1, 0.24 $\pm$ 0.02) ~hr$^{-1}$, respectively, and the long-time extrapolated path change $\delta L_{\infty}$ are 244 $\pm$ 80, 114 $\pm$ 3, -50 $\pm$ 10, and -198 $\pm$ 12 nm, respectively. (b) Variation in the temperature as the interferometer is heated. The rate constants for the exponential fit for intervals 1-4 are 1.101 $\pm$ 0.002, 1.330 $\pm$ 0.001, 1.277 $\pm$ 0.003 and 1.48 $\pm$ 0.01~hr$^{-1}$, respectively and the extrapolated temperature is 27.89 $\pm$ 0.01, 34.69 $\pm$ 0.01, 43.90 $\pm$ 0.08, 50.94 $\pm$ 0.04~$^{\circ}$C, respectively. (c) Temperature dependence of the long-time extrapolated path change along with a fit to a quadratic function.}
	\label{TDFS2}
\end{figure}

We observe similar behavior for the 1.25~GHz interferometer as shown in Fig.~\ref{TDFS2}. Importantly, we observe that the contribution to $\delta L$ from the glass can counteract that due to the aluminum housing for the last three intervals. Using the same fitting procedure as used above, we find that the average rate constant for the temperature change is 1.297$\pm$ 0.003 hr$^{-1}$ and the two rate constants for the path change are 1.57$\pm$ 0.08 hr$^{-1}$ and 0.33$\pm$ 0.06 hr$^{-1}$. Again, there is a strong correlation between the temperature rate constant and the fast rate constant for the path change, indicating that the aluminum housing is playing an important role in our observations. Using our fit to extrapolate to long times for each interval, we find that the path change is a nonlinear function of $T$ (well fit by a quadratic in this case) and hence a single value for the TDPS does not adequately characterize this device. Just considering the data point for the first interval, the inferred TDPS is 50 $\pm$ 17 ~nm/$^{\circ}$C, which again exceeds the specification of 22~nm/$^{\circ}$C. However, the TDPS is zero at $37.1~^{\circ}$C based on our fit to a quadratic.

\section{Application in Four-Dimensional time-frequency QKD}\label{FourDimensionalQKD}
\begin{figure}[htp]
	\begin{center}
		\includegraphics[width = 0.5\textwidth]{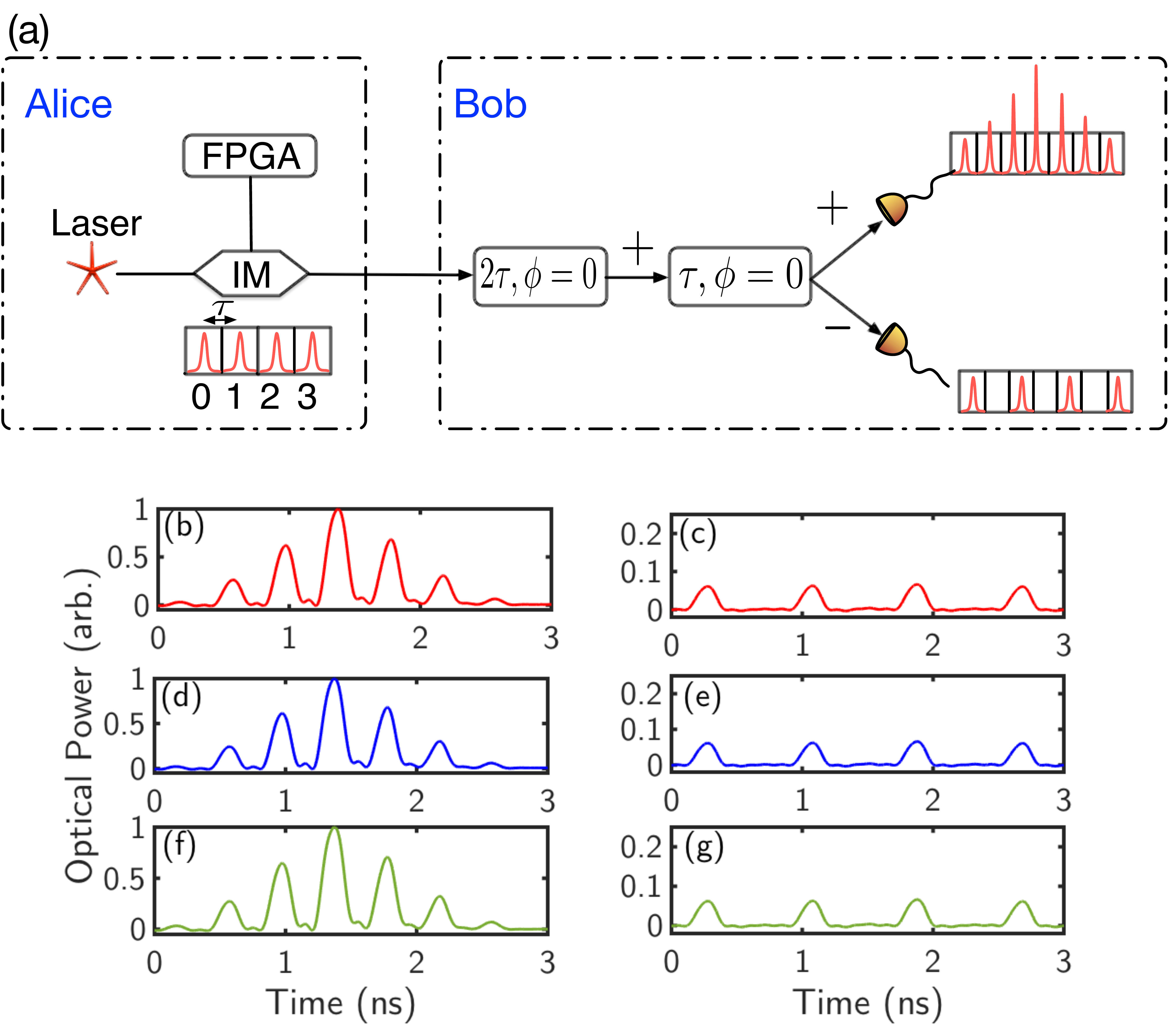}
		\caption{(a) Schematic illustrating the experimental details for creating and detecting $|\Psi_{f_0}\rangle$ in $d = 4$. A 1550 nm continuous wave laser (Agilent HP81862A, power 1 mW) is modulated into a four-peaked wavepacket (100~ps pulse-width in 400~ps time bins) using a Mach-Zehnder modulator (EOSPACE AX-0K1-12-PFA-PFAP-R3-SRF1W). The optical modulator driver (JDSU H301), used as an amplifier at the input of the modulator, is driven with a field-programmable gate array (Stratix V FPGA 5SGXEA7N2F40C2N, Terasic Stratix V Signal Integrity Kit). The signals are passed through 2 interferometers of time-delays 800 ps and 400 ps, respectively and detected using two-high speed photoreceivers (Miteq DR-125G, 12.5~GHz bandwidth) and an oscilloscope (Agilent Infiniium DSO90804A). (b) and (c) The constructive and destructive interference patterns at the start of the measurement. (d) and (e) ((f) and (g)) show the interference 30~(60)~minutes after the start of the data collection.}
		\label{ExperimentalSetup}
	\end{center}
\end{figure}

To demonstrate the applicability of these interferometers in high-dimensional ($d=4$) time-frequency QKD, we use them to measure the classical analog of the frequency state $|\Psi_{f_0}\rangle$ with the setup shown in Fig.~\ref{ExperimentalSetup}(a) {\em i. e.}, the equivalent state generated with strong coherent light and detected using an InGaAs pin photodiode. These observations are directly applicable to a QKD system in that the quantum signal states can be generated by strongly attenuating our classical source and photon counting detectors replacing the photodiode. Thus, the observed interference pattern with weak and strong coherent states should be identical and our results in the classical regime should be directly applicable in the single-photon regime.

Pulses are created from a continuous-wave laser beam using a Mach-Zehnder modulator in the form of frames with four distinct peaks of width 100~ps. The optical modulator is driven by an arbitrary serial pattern generator produced by a 10 GHz transceiver from a field-programmable gate array, amplified with a driver amplifier. The time-bin width is set to 400~ps as shown in Fig.~\ref{ExperimentalSetup}(a), matched to the time-delay $\tau$ of the second interferometer (2.5 GHz FSR) in the cascade (note that the other interferometer with path difference $\tau$ is not shown for clarity). Over the short duration of a frame, the phase of the laser is constant, and hence all the pulses within a frame have the same relative phase. The outputs of the interferometers are measured using two high-speed photoreceivers and an oscilloscope with a bandwidth of 8 GHz. We note that the bandwidth of the oscilloscope is smaller than what is required to measure 10 GHz short pulses and hence introduces a systematic uncertainty in the measurement. Specifically, the calculation of the visibility is affected by our measurement apparatus because the observed interference peaks are broadened. The observed constructive (destructive) interference pattern is shown in Fig.~\ref{ExperimentalSetup}(b) (Fig.~\ref{ExperimentalSetup}(c)) similar to the expected intensity pattern as shown in Fig.~\ref{FrequencyBasisMeasurement}.

The visibility of the interference pattern is calculated using Eq.~\ref{Vis}, where $\mathcal{P}_+$ is the area under the constructive interference peak, and $\mathcal{P}_-$ is the area under destructive interference peak. The central peak with the largest amplitude in `$+$' output is the constructive interference peak, while the corresponding time-bin in the `$-$' output is the destructive interference peak. Based on this, we find that $\mathcal{V}>$ 99\% over an hour (Fig.~\ref{ExperimentalSetup}(d, e, f, g)) in the nominally constant laboratory environment where the temperature changes no more than $\pm0.1^{\circ}$C, similar to what we observe in Sec.~\ref{InterferometerPerformance} above using a continuous-wave laser.

\section{Conclusion}
In conclusion, we demonstrate high environmental stability for commercially-available temperature-compensated time-delay interferometers for application in discrete-variable time-frequency QKD. In particular, we observe that both the 2.5~GHz and 1.25~GHz interferometers have a path-length stability of better than 3~nm when the temperature is maintained within $\pm 0.1^{\circ}$C in a laboratory environment. In addition, when heated in a controlled manner, we observe a temperature dependent path-length shift (TDPS) of 26 $\pm$ 9~nm/$^{\circ}$C for the 2.5~GHz device. For the 1.25~GHz interferometer, we observe a nonlinear change in path-length as a function of temperature, which is locally at zero at 37.1~$^{\circ}$C, and 50$\pm$ 17~nm/$^{\circ}$C at 22$^{\circ}$C. We argue that for such nonlinear devices, the TDPS metric is not sufficient. Rather, we have to assess both the stability at a constant temperature over a long time scale, as well as over a wide range of temperature. We also investigate the possibility of using these passive interferometers as components in time-bin encoding QKD. We observe a maximum change of visibility of less than 1.0 \% over an hour, which shows that if the temperature of these devices is maintained actively within $\pm 0.1 ^{\circ}$C, then they can indeed be used for long distance high-dimensional time-bin encoding QKD. We also demonstrate a proof-of-principle experiment with classical coherent pulses that show predictable interference pattern with high visibility. We conclude that these devices are suitable for realizing a high-dimension discrete-variable time-frequency QKD system.

\section{Acknowledgment}
We gratefully acknowledge discussions of this work with Paul Kwiat, Charles Ci Wen Lim, Bill Brown, Clinton Cahall, and Ludovic Fulop, and the financial support of the ONR MURI program on Wavelength-Agile Quantum Key Distribution in a Marine Environment, Grant \# N00014-13-1-0627. 

\bigskip
\bibliography{sample}

\begin{thebibliography}{40}%
\makeatletter
\providecommand \@ifxundefined [1]{%
 \@ifx{#1\undefined}
}%
\providecommand \@ifnum [1]{%
 \ifnum #1\expandafter \@firstoftwo
 \else \expandafter \@secondoftwo
 \fi
}%
\providecommand \@ifx [1]{%
 \ifx #1\expandafter \@firstoftwo
 \else \expandafter \@secondoftwo
 \fi
}%
\providecommand \natexlab [1]{#1}%
\providecommand \enquote  [1]{``#1''}%
\providecommand \bibnamefont  [1]{#1}%
\providecommand \bibfnamefont [1]{#1}%
\providecommand \citenamefont [1]{#1}%
\providecommand \href@noop [0]{\@secondoftwo}%
\providecommand \href [0]{\begingroup \@sanitize@url \@href}%
\providecommand \@href[1]{\@@startlink{#1}\@@href}%
\providecommand \@@href[1]{\endgroup#1\@@endlink}%
\providecommand \@sanitize@url [0]{\catcode `\\12\catcode `\$12\catcode
  `\&12\catcode `\#12\catcode `\^12\catcode `\_12\catcode `\%12\relax}%
\providecommand \@@startlink[1]{}%
\providecommand \@@endlink[0]{}%
\providecommand \url  [0]{\begingroup\@sanitize@url \@url }%
\providecommand \@url [1]{\endgroup\@href {#1}{\urlprefix }}%
\providecommand \urlprefix  [0]{URL }%
\providecommand \Eprint [0]{\href }%
\providecommand \doibase [0]{http://dx.doi.org/}%
\providecommand \selectlanguage [0]{\@gobble}%
\providecommand \bibinfo  [0]{\@secondoftwo}%
\providecommand \bibfield  [0]{\@secondoftwo}%
\providecommand \translation [1]{[#1]}%
\providecommand \BibitemOpen [0]{}%
\providecommand \bibitemStop [0]{}%
\providecommand \bibitemNoStop [0]{.\EOS\space}%
\providecommand \EOS [0]{\spacefactor3000\relax}%
\providecommand \BibitemShut  [1]{\csname bibitem#1\endcsname}%
\let\auto@bib@innerbib\@empty
\bibitem [{\citenamefont {Barnett}(2009)}]{BarnettBook}%
  \BibitemOpen
  \bibfield  {author} {\bibinfo {author} {\bibfnamefont {S.}~\bibnamefont
  {Barnett}},\ }\href {https://books.google.com/books?id=A2k4HH2tFR8C} {\emph
  {\bibinfo {title} {Quantum Information}}},\ Oxford Master Series in Physics\
  (\bibinfo  {publisher} {OUP Oxford},\ \bibinfo {year} {2009})\BibitemShut
  {NoStop}%
\bibitem [{\citenamefont {Walenta}\ \emph {et~al.}(2014)\citenamefont
  {Walenta}, \citenamefont {Burg}, \citenamefont {Caselunghe}, \citenamefont
  {Constantin}, \citenamefont {Gisin}, \citenamefont {Guinnard}, \citenamefont
  {Houlmann}, \citenamefont {Junod}, \citenamefont {Korzh}, \citenamefont
  {Kulesza}, \citenamefont {Legré}, \citenamefont {Lim}, \citenamefont
  {Lunghi}, \citenamefont {Monat}, \citenamefont {Portmann}, \citenamefont
  {Soucarros}, \citenamefont {Thew}, \citenamefont {Trinkler}, \citenamefont
  {Trolliet}, \citenamefont {Vannel},\ and\ \citenamefont
  {Zbinden}}]{Walenta14}%
  \BibitemOpen
  \bibfield  {author} {\bibinfo {author} {\bibfnamefont {N.}~\bibnamefont
  {Walenta}}, \bibinfo {author} {\bibfnamefont {A.}~\bibnamefont {Burg}},
  \bibinfo {author} {\bibfnamefont {D.}~\bibnamefont {Caselunghe}}, \bibinfo
  {author} {\bibfnamefont {J.}~\bibnamefont {Constantin}}, \bibinfo {author}
  {\bibfnamefont {N.}~\bibnamefont {Gisin}}, \bibinfo {author} {\bibfnamefont
  {O.}~\bibnamefont {Guinnard}}, \bibinfo {author} {\bibfnamefont
  {R.}~\bibnamefont {Houlmann}}, \bibinfo {author} {\bibfnamefont
  {P.}~\bibnamefont {Junod}}, \bibinfo {author} {\bibfnamefont
  {B.}~\bibnamefont {Korzh}}, \bibinfo {author} {\bibfnamefont
  {N.}~\bibnamefont {Kulesza}}, \bibinfo {author} {\bibfnamefont
  {M.}~\bibnamefont {Legré}}, \bibinfo {author} {\bibfnamefont {C.~W.}\
  \bibnamefont {Lim}}, \bibinfo {author} {\bibfnamefont {T.}~\bibnamefont
  {Lunghi}}, \bibinfo {author} {\bibfnamefont {L.}~\bibnamefont {Monat}},
  \bibinfo {author} {\bibfnamefont {C.}~\bibnamefont {Portmann}}, \bibinfo
  {author} {\bibfnamefont {M.}~\bibnamefont {Soucarros}}, \bibinfo {author}
  {\bibfnamefont {R.~T.}\ \bibnamefont {Thew}}, \bibinfo {author}
  {\bibfnamefont {P.}~\bibnamefont {Trinkler}}, \bibinfo {author}
  {\bibfnamefont {G.}~\bibnamefont {Trolliet}}, \bibinfo {author}
  {\bibfnamefont {F.}~\bibnamefont {Vannel}}, \ and\ \bibinfo {author}
  {\bibfnamefont {H.}~\bibnamefont {Zbinden}},\ }\href
  {http://stacks.iop.org/1367-2630/16/i=1/a=013047} {\bibfield  {journal}
  {\bibinfo  {journal} {New Journal of Physics}\ }\textbf {\bibinfo {volume}
  {16}},\ \bibinfo {pages} {013047} (\bibinfo {year} {2014})}\BibitemShut
  {NoStop}%
\bibitem [{\citenamefont {Lucamarini}\ \emph {et~al.}(2013)\citenamefont
  {Lucamarini}, \citenamefont {Patel}, \citenamefont {Dynes}, \citenamefont
  {Fr\"{o}hlich}, \citenamefont {Sharpe}, \citenamefont {Dixon}, \citenamefont
  {Yuan}, \citenamefont {Penty},\ and\ \citenamefont {Shields}}]{Lucamarini13}%
  \BibitemOpen
  \bibfield  {author} {\bibinfo {author} {\bibfnamefont {M.}~\bibnamefont
  {Lucamarini}}, \bibinfo {author} {\bibfnamefont {K.~A.}\ \bibnamefont
  {Patel}}, \bibinfo {author} {\bibfnamefont {J.~F.}\ \bibnamefont {Dynes}},
  \bibinfo {author} {\bibfnamefont {B.}~\bibnamefont {Fr\"{o}hlich}}, \bibinfo
  {author} {\bibfnamefont {A.~W.}\ \bibnamefont {Sharpe}}, \bibinfo {author}
  {\bibfnamefont {A.~R.}\ \bibnamefont {Dixon}}, \bibinfo {author}
  {\bibfnamefont {Z.~L.}\ \bibnamefont {Yuan}}, \bibinfo {author}
  {\bibfnamefont {R.~V.}\ \bibnamefont {Penty}}, \ and\ \bibinfo {author}
  {\bibfnamefont {A.~J.}\ \bibnamefont {Shields}},\ }\href {\doibase
  10.1364/OE.21.024550} {\bibfield  {journal} {\bibinfo  {journal} {Opt.
  Express}\ }\textbf {\bibinfo {volume} {21}},\ \bibinfo {pages} {24550}
  (\bibinfo {year} {2013})}\BibitemShut {NoStop}%
\bibitem [{\citenamefont {Lim}\ \emph {et~al.}(2014)\citenamefont {Lim},
  \citenamefont {Curty}, \citenamefont {Walenta}, \citenamefont {Xu},\ and\
  \citenamefont {Zbinden}}]{CharlesDecoy}%
  \BibitemOpen
  \bibfield  {author} {\bibinfo {author} {\bibfnamefont {C.~C.~W.}\
  \bibnamefont {Lim}}, \bibinfo {author} {\bibfnamefont {M.}~\bibnamefont
  {Curty}}, \bibinfo {author} {\bibfnamefont {N.}~\bibnamefont {Walenta}},
  \bibinfo {author} {\bibfnamefont {F.}~\bibnamefont {Xu}}, \ and\ \bibinfo
  {author} {\bibfnamefont {H.}~\bibnamefont {Zbinden}},\ }\href {\doibase
  10.1103/PhysRevA.89.022307} {\bibfield  {journal} {\bibinfo  {journal} {Phys.
  Rev. A}\ }\textbf {\bibinfo {volume} {89}},\ \bibinfo {pages} {022307}
  (\bibinfo {year} {2014})}\BibitemShut {NoStop}%
\bibitem [{\citenamefont {Korzh}\ \emph {et~al.}(2015)\citenamefont {Korzh},
  \citenamefont {Lim}, \citenamefont {Houlmann}, \citenamefont {Gisin},
  \citenamefont {Li}, \citenamefont {Nolan}, \citenamefont {Sanguinetti},
  \citenamefont {Thew},\ and\ \citenamefont {Zbinden}}]{Boris2014}%
  \BibitemOpen
  \bibfield  {author} {\bibinfo {author} {\bibfnamefont {B.}~\bibnamefont
  {Korzh}}, \bibinfo {author} {\bibfnamefont {C.~C.~W.}\ \bibnamefont {Lim}},
  \bibinfo {author} {\bibfnamefont {R.}~\bibnamefont {Houlmann}}, \bibinfo
  {author} {\bibfnamefont {N.}~\bibnamefont {Gisin}}, \bibinfo {author}
  {\bibfnamefont {M.~J.}\ \bibnamefont {Li}}, \bibinfo {author} {\bibfnamefont
  {D.}~\bibnamefont {Nolan}}, \bibinfo {author} {\bibfnamefont
  {B.}~\bibnamefont {Sanguinetti}}, \bibinfo {author} {\bibfnamefont
  {R.}~\bibnamefont {Thew}}, \ and\ \bibinfo {author} {\bibfnamefont
  {H.}~\bibnamefont {Zbinden}},\ }\href {\doibase 10.1038/nphoton.2014.327}
  {\bibfield  {journal} {\bibinfo  {journal} {Nature Photonics}\ }\textbf
  {\bibinfo {volume} {9}},\ \bibinfo {pages} {163–168} (\bibinfo {year}
  {2015})}\BibitemShut {NoStop}%
\bibitem [{\citenamefont {Yin}\ \emph {et~al.}(2016)\citenamefont {Yin},
  \citenamefont {Chen}, \citenamefont {Yu}, \citenamefont {Liu}, \citenamefont
  {You}, \citenamefont {Zhou}, \citenamefont {Chen}, \citenamefont {Mao},
  \citenamefont {Huang}, \citenamefont {Zhang} \emph {et~al.}}]{Yin2016}%
  \BibitemOpen
  \bibfield  {author} {\bibinfo {author} {\bibfnamefont {H.-L.}\ \bibnamefont
  {Yin}}, \bibinfo {author} {\bibfnamefont {T.-Y.}\ \bibnamefont {Chen}},
  \bibinfo {author} {\bibfnamefont {Z.-W.}\ \bibnamefont {Yu}}, \bibinfo
  {author} {\bibfnamefont {H.}~\bibnamefont {Liu}}, \bibinfo {author}
  {\bibfnamefont {L.-X.}\ \bibnamefont {You}}, \bibinfo {author} {\bibfnamefont
  {Y.-H.}\ \bibnamefont {Zhou}}, \bibinfo {author} {\bibfnamefont {S.-J.}\
  \bibnamefont {Chen}}, \bibinfo {author} {\bibfnamefont {Y.}~\bibnamefont
  {Mao}}, \bibinfo {author} {\bibfnamefont {M.-Q.}\ \bibnamefont {Huang}},
  \bibinfo {author} {\bibfnamefont {W.-J.}\ \bibnamefont {Zhang}},  \emph
  {et~al.},\ }\href@noop {} {\bibfield  {journal} {\bibinfo  {journal} {arXiv
  preprint arXiv:1606.06821}\ } (\bibinfo {year} {2016})}\BibitemShut {NoStop}%
\bibitem [{\citenamefont {Bechmann-Pasquinucci}\ and\ \citenamefont
  {Tittel}(2000)}]{Bechmann01}%
  \BibitemOpen
  \bibfield  {author} {\bibinfo {author} {\bibfnamefont {H.}~\bibnamefont
  {Bechmann-Pasquinucci}}\ and\ \bibinfo {author} {\bibfnamefont
  {W.}~\bibnamefont {Tittel}},\ }\href {\doibase 10.1103/PhysRevA.61.062308}
  {\bibfield  {journal} {\bibinfo  {journal} {Phys. Rev. A}\ }\textbf {\bibinfo
  {volume} {61}},\ \bibinfo {pages} {062308} (\bibinfo {year}
  {2000})}\BibitemShut {NoStop}%
\bibitem [{\citenamefont {Cerf}\ \emph {et~al.}(2002)\citenamefont {Cerf},
  \citenamefont {Bourennane}, \citenamefont {Karlsson},\ and\ \citenamefont
  {Gisin}}]{Cerf02}%
  \BibitemOpen
  \bibfield  {author} {\bibinfo {author} {\bibfnamefont {N.~J.}\ \bibnamefont
  {Cerf}}, \bibinfo {author} {\bibfnamefont {M.}~\bibnamefont {Bourennane}},
  \bibinfo {author} {\bibfnamefont {A.}~\bibnamefont {Karlsson}}, \ and\
  \bibinfo {author} {\bibfnamefont {N.}~\bibnamefont {Gisin}},\ }\href
  {\doibase 10.1103/PhysRevLett.88.127902} {\bibfield  {journal} {\bibinfo
  {journal} {Phys. Rev. Lett.}\ }\textbf {\bibinfo {volume} {88}},\ \bibinfo
  {pages} {127902} (\bibinfo {year} {2002})}\BibitemShut {NoStop}%
\bibitem [{\citenamefont {Scarani}\ \emph {et~al.}(2009)\citenamefont
  {Scarani}, \citenamefont {Bechmann-Pasquinucci}, \citenamefont {Cerf},
  \citenamefont {Du\ifmmode~\check{s}\else \v{s}\fi{}ek}, \citenamefont
  {L\"utkenhaus},\ and\ \citenamefont {Peev}}]{Valero09}%
  \BibitemOpen
  \bibfield  {author} {\bibinfo {author} {\bibfnamefont {V.}~\bibnamefont
  {Scarani}}, \bibinfo {author} {\bibfnamefont {H.}~\bibnamefont
  {Bechmann-Pasquinucci}}, \bibinfo {author} {\bibfnamefont {N.~J.}\
  \bibnamefont {Cerf}}, \bibinfo {author} {\bibfnamefont {M.}~\bibnamefont
  {Du\ifmmode~\check{s}\else \v{s}\fi{}ek}}, \bibinfo {author} {\bibfnamefont
  {N.}~\bibnamefont {L\"utkenhaus}}, \ and\ \bibinfo {author} {\bibfnamefont
  {M.}~\bibnamefont {Peev}},\ }\href {\doibase 10.1103/RevModPhys.81.1301}
  {\bibfield  {journal} {\bibinfo  {journal} {Rev. Mod. Phys.}\ }\textbf
  {\bibinfo {volume} {81}},\ \bibinfo {pages} {1301} (\bibinfo {year}
  {2009})}\BibitemShut {NoStop}%
\bibitem [{\citenamefont {Sheridan}\ and\ \citenamefont
  {Scarani}(2010)}]{Valero10}%
  \BibitemOpen
  \bibfield  {author} {\bibinfo {author} {\bibfnamefont {L.}~\bibnamefont
  {Sheridan}}\ and\ \bibinfo {author} {\bibfnamefont {V.}~\bibnamefont
  {Scarani}},\ }\href {\doibase 10.1103/PhysRevA.82.030301} {\bibfield
  {journal} {\bibinfo  {journal} {Phys. Rev. A}\ }\textbf {\bibinfo {volume}
  {82}},\ \bibinfo {pages} {030301} (\bibinfo {year} {2010})}\BibitemShut
  {NoStop}%
\bibitem [{\citenamefont {Brougham}\ \emph {et~al.}(2013)\citenamefont
  {Brougham}, \citenamefont {Barnett}, \citenamefont {McCusker}, \citenamefont
  {Kwiat},\ and\ \citenamefont {Gauthier}}]{Brougham13}%
  \BibitemOpen
  \bibfield  {author} {\bibinfo {author} {\bibfnamefont {T.}~\bibnamefont
  {Brougham}}, \bibinfo {author} {\bibfnamefont {S.~M.}\ \bibnamefont
  {Barnett}}, \bibinfo {author} {\bibfnamefont {K.~T.}\ \bibnamefont
  {McCusker}}, \bibinfo {author} {\bibfnamefont {P.~G.}\ \bibnamefont {Kwiat}},
  \ and\ \bibinfo {author} {\bibfnamefont {D.~J.}\ \bibnamefont {Gauthier}},\
  }\href {http://stacks.iop.org/0953-4075/46/i=10/a=104010} {\bibfield
  {journal} {\bibinfo  {journal} {Journal of Physics B: Atomic, Molecular and
  Optical Physics}\ }\textbf {\bibinfo {volume} {46}},\ \bibinfo {pages}
  {104010} (\bibinfo {year} {2013})}\BibitemShut {NoStop}%
\bibitem [{\citenamefont {Mower}\ \emph {et~al.}(2013)\citenamefont {Mower},
  \citenamefont {Zhang}, \citenamefont {Desjardins}, \citenamefont {Lee},
  \citenamefont {Shapiro},\ and\ \citenamefont {Englund}}]{Shapiro13}%
  \BibitemOpen
  \bibfield  {author} {\bibinfo {author} {\bibfnamefont {J.}~\bibnamefont
  {Mower}}, \bibinfo {author} {\bibfnamefont {Z.}~\bibnamefont {Zhang}},
  \bibinfo {author} {\bibfnamefont {P.}~\bibnamefont {Desjardins}}, \bibinfo
  {author} {\bibfnamefont {C.}~\bibnamefont {Lee}}, \bibinfo {author}
  {\bibfnamefont {J.~H.}\ \bibnamefont {Shapiro}}, \ and\ \bibinfo {author}
  {\bibfnamefont {D.}~\bibnamefont {Englund}},\ }\href {\doibase
  10.1103/PhysRevA.87.062322} {\bibfield  {journal} {\bibinfo  {journal} {Phys.
  Rev. A}\ }\textbf {\bibinfo {volume} {87}},\ \bibinfo {pages} {062322}
  (\bibinfo {year} {2013})}\BibitemShut {NoStop}%
\bibitem [{\citenamefont {Nunn}\ \emph {et~al.}(2013)\citenamefont {Nunn},
  \citenamefont {Wright}, \citenamefont {S\"{o}ller}, \citenamefont {Zhang},
  \citenamefont {Walmsley},\ and\ \citenamefont {Smith}}]{Nunn13}%
  \BibitemOpen
  \bibfield  {author} {\bibinfo {author} {\bibfnamefont {J.}~\bibnamefont
  {Nunn}}, \bibinfo {author} {\bibfnamefont {L.~J.}\ \bibnamefont {Wright}},
  \bibinfo {author} {\bibfnamefont {C.}~\bibnamefont {S\"{o}ller}}, \bibinfo
  {author} {\bibfnamefont {L.}~\bibnamefont {Zhang}}, \bibinfo {author}
  {\bibfnamefont {I.~A.}\ \bibnamefont {Walmsley}}, \ and\ \bibinfo {author}
  {\bibfnamefont {B.~J.}\ \bibnamefont {Smith}},\ }\href {\doibase
  10.1364/OE.21.015959} {\bibfield  {journal} {\bibinfo  {journal} {Opt.
  Express}\ }\textbf {\bibinfo {volume} {21}},\ \bibinfo {pages} {15959}
  (\bibinfo {year} {2013})}\BibitemShut {NoStop}%
\bibitem [{\citenamefont {Zhang}\ \emph {et~al.}(2014)\citenamefont {Zhang},
  \citenamefont {Mower}, \citenamefont {Englund}, \citenamefont {Wong},\ and\
  \citenamefont {Shapiro}}]{Shapiro14}%
  \BibitemOpen
  \bibfield  {author} {\bibinfo {author} {\bibfnamefont {Z.}~\bibnamefont
  {Zhang}}, \bibinfo {author} {\bibfnamefont {J.}~\bibnamefont {Mower}},
  \bibinfo {author} {\bibfnamefont {D.}~\bibnamefont {Englund}}, \bibinfo
  {author} {\bibfnamefont {F.~N.~C.}\ \bibnamefont {Wong}}, \ and\ \bibinfo
  {author} {\bibfnamefont {J.~H.}\ \bibnamefont {Shapiro}},\ }\href {\doibase
  10.1103/PhysRevLett.112.120506} {\bibfield  {journal} {\bibinfo  {journal}
  {Phys. Rev. Lett.}\ }\textbf {\bibinfo {volume} {112}},\ \bibinfo {pages}
  {120506} (\bibinfo {year} {2014})}\BibitemShut {NoStop}%
\bibitem [{\citenamefont {Gauthier}\ \emph {et~al.}(2013)\citenamefont
  {Gauthier}, \citenamefont {Wildfeuer}, \citenamefont {Guilbert},
  \citenamefont {Stipcevic}, \citenamefont {Christensen}, \citenamefont
  {Kumor}, \citenamefont {Kwiat}, \citenamefont {McCusker}, \citenamefont
  {Brougham},\ and\ \citenamefont {Barnett}}]{Dan14}%
  \BibitemOpen
  \bibfield  {author} {\bibinfo {author} {\bibfnamefont {D.~J.}\ \bibnamefont
  {Gauthier}}, \bibinfo {author} {\bibfnamefont {C.~F.}\ \bibnamefont
  {Wildfeuer}}, \bibinfo {author} {\bibfnamefont {H.}~\bibnamefont {Guilbert}},
  \bibinfo {author} {\bibfnamefont {M.}~\bibnamefont {Stipcevic}}, \bibinfo
  {author} {\bibfnamefont {B.~G.}\ \bibnamefont {Christensen}}, \bibinfo
  {author} {\bibfnamefont {D.}~\bibnamefont {Kumor}}, \bibinfo {author}
  {\bibfnamefont {P.}~\bibnamefont {Kwiat}}, \bibinfo {author} {\bibfnamefont
  {K.~T.}\ \bibnamefont {McCusker}}, \bibinfo {author} {\bibfnamefont
  {T.}~\bibnamefont {Brougham}}, \ and\ \bibinfo {author} {\bibfnamefont
  {S.}~\bibnamefont {Barnett}},\ }in\ \href {\doibase 10.1364/QIM.2013.W2A.2}
  {\emph {\bibinfo {booktitle} {The Rochester Conferences on Coherence and
  Quantum Optics and the Quantum Information and Measurement meeting}}}\
  (\bibinfo  {publisher} {Optical Society of America},\ \bibinfo {year}
  {2013})\ p.\ \bibinfo {pages} {W2A.2}\BibitemShut {NoStop}%
\bibitem [{\citenamefont {Zhong}\ \emph {et~al.}(2015)\citenamefont {Zhong},
  \citenamefont {Zhou}, \citenamefont {Horansky}, \citenamefont {Lee},
  \citenamefont {Verma}, \citenamefont {Lita}, \citenamefont {Restelli},
  \citenamefont {Bienfang}, \citenamefont {Mirin}, \citenamefont {Gerrits},
  \citenamefont {Nam}, \citenamefont {Marsili}, \citenamefont {Shaw},
  \citenamefont {Zhang}, \citenamefont {Wang}, \citenamefont {Englund},
  \citenamefont {Wornell}, \citenamefont {Shapiro},\ and\ \citenamefont
  {Wong}}]{MIT2015}%
  \BibitemOpen
  \bibfield  {author} {\bibinfo {author} {\bibfnamefont {T.}~\bibnamefont
  {Zhong}}, \bibinfo {author} {\bibfnamefont {H.}~\bibnamefont {Zhou}},
  \bibinfo {author} {\bibfnamefont {R.~D.}\ \bibnamefont {Horansky}}, \bibinfo
  {author} {\bibfnamefont {C.}~\bibnamefont {Lee}}, \bibinfo {author}
  {\bibfnamefont {V.~B.}\ \bibnamefont {Verma}}, \bibinfo {author}
  {\bibfnamefont {A.~E.}\ \bibnamefont {Lita}}, \bibinfo {author}
  {\bibfnamefont {A.}~\bibnamefont {Restelli}}, \bibinfo {author}
  {\bibfnamefont {J.~C.}\ \bibnamefont {Bienfang}}, \bibinfo {author}
  {\bibfnamefont {R.~P.}\ \bibnamefont {Mirin}}, \bibinfo {author}
  {\bibfnamefont {T.}~\bibnamefont {Gerrits}}, \bibinfo {author} {\bibfnamefont
  {S.~W.}\ \bibnamefont {Nam}}, \bibinfo {author} {\bibfnamefont
  {F.}~\bibnamefont {Marsili}}, \bibinfo {author} {\bibfnamefont {M.~D.}\
  \bibnamefont {Shaw}}, \bibinfo {author} {\bibfnamefont {Z.}~\bibnamefont
  {Zhang}}, \bibinfo {author} {\bibfnamefont {L.}~\bibnamefont {Wang}},
  \bibinfo {author} {\bibfnamefont {D.}~\bibnamefont {Englund}}, \bibinfo
  {author} {\bibfnamefont {G.~W.}\ \bibnamefont {Wornell}}, \bibinfo {author}
  {\bibfnamefont {J.~H.}\ \bibnamefont {Shapiro}}, \ and\ \bibinfo {author}
  {\bibfnamefont {F.~N.~C.}\ \bibnamefont {Wong}},\ }\href
  {http://stacks.iop.org/1367-2630/17/i=2/a=022002} {\bibfield  {journal}
  {\bibinfo  {journal} {New Journal of Physics}\ }\textbf {\bibinfo {volume}
  {17}},\ \bibinfo {pages} {022002} (\bibinfo {year} {2015})}\BibitemShut
  {NoStop}%
\bibitem [{\citenamefont {Brougham}\ \emph {et~al.}(2015)\citenamefont
  {Brougham}, \citenamefont {Wildfeuer}, \citenamefont {Barnett},\ and\
  \citenamefont {Gauthier}}]{Brougham15}%
  \BibitemOpen
  \bibfield  {author} {\bibinfo {author} {\bibfnamefont {T.}~\bibnamefont
  {Brougham}}, \bibinfo {author} {\bibfnamefont {C.~F.}\ \bibnamefont
  {Wildfeuer}}, \bibinfo {author} {\bibfnamefont {S.~M.}\ \bibnamefont
  {Barnett}}, \ and\ \bibinfo {author} {\bibfnamefont {D.~J.}\ \bibnamefont
  {Gauthier}},\ }\href@noop {} {\enquote {\bibinfo {title} {The information of
  high-dimensional time-bin encoded photons},}\ } (\bibinfo {year} {2015}),\
  \Eprint {http://arxiv.org/abs/arXiv:1506.04420} {arXiv:1506.04420}
  \BibitemShut {NoStop}%
\bibitem [{\citenamefont {Brecht}\ \emph {et~al.}(2015)\citenamefont {Brecht},
  \citenamefont {Reddy}, \citenamefont {Silberhorn},\ and\ \citenamefont
  {Raymer}}]{Brecht15}%
  \BibitemOpen
  \bibfield  {author} {\bibinfo {author} {\bibfnamefont {B.}~\bibnamefont
  {Brecht}}, \bibinfo {author} {\bibfnamefont {D.~V.}\ \bibnamefont {Reddy}},
  \bibinfo {author} {\bibfnamefont {C.}~\bibnamefont {Silberhorn}}, \ and\
  \bibinfo {author} {\bibfnamefont {M.~G.}\ \bibnamefont {Raymer}},\ }\href
  {\doibase 10.1103/PhysRevX.5.041017} {\bibfield  {journal} {\bibinfo
  {journal} {Phys. Rev. X}\ }\textbf {\bibinfo {volume} {5}},\ \bibinfo {pages}
  {041017} (\bibinfo {year} {2015})}\BibitemShut {NoStop}%
\bibitem [{\citenamefont {Mirhosseini}\ \emph {et~al.}(2015)\citenamefont
  {Mirhosseini}, \citenamefont {Magaña-Loaiza}, \citenamefont {O’Sullivan},
  \citenamefont {Rodenburg}, \citenamefont {Malik}, \citenamefont {Lavery},
  \citenamefont {Padgett}, \citenamefont {Gauthier},\ and\ \citenamefont
  {Boyd}}]{Boyd14}%
  \BibitemOpen
  \bibfield  {author} {\bibinfo {author} {\bibfnamefont {M.}~\bibnamefont
  {Mirhosseini}}, \bibinfo {author} {\bibfnamefont {O.~S.}\ \bibnamefont
  {Magaña-Loaiza}}, \bibinfo {author} {\bibfnamefont {M.~N.}\ \bibnamefont
  {O’Sullivan}}, \bibinfo {author} {\bibfnamefont {B.}~\bibnamefont
  {Rodenburg}}, \bibinfo {author} {\bibfnamefont {M.}~\bibnamefont {Malik}},
  \bibinfo {author} {\bibfnamefont {M.~P.~J.}\ \bibnamefont {Lavery}}, \bibinfo
  {author} {\bibfnamefont {M.~J.}\ \bibnamefont {Padgett}}, \bibinfo {author}
  {\bibfnamefont {D.~J.}\ \bibnamefont {Gauthier}}, \ and\ \bibinfo {author}
  {\bibfnamefont {R.~W.}\ \bibnamefont {Boyd}},\ }\href
  {http://stacks.iop.org/1367-2630/17/i=3/a=033033} {\bibfield  {journal}
  {\bibinfo  {journal} {New Journal of Physics}\ }\textbf {\bibinfo {volume}
  {17}},\ \bibinfo {pages} {033033} (\bibinfo {year} {2015})}\BibitemShut
  {NoStop}%
\bibitem [{\citenamefont {Barreiro}\ \emph
  {et~al.}(2005{\natexlab{a}})\citenamefont {Barreiro}, \citenamefont
  {Langford}, \citenamefont {Peters},\ and\ \citenamefont {Kwiat}}]{Kwiat05}%
  \BibitemOpen
  \bibfield  {author} {\bibinfo {author} {\bibfnamefont {J.~T.}\ \bibnamefont
  {Barreiro}}, \bibinfo {author} {\bibfnamefont {N.~K.}\ \bibnamefont
  {Langford}}, \bibinfo {author} {\bibfnamefont {N.~A.}\ \bibnamefont
  {Peters}}, \ and\ \bibinfo {author} {\bibfnamefont {P.~G.}\ \bibnamefont
  {Kwiat}},\ }\href {\doibase 10.1103/PhysRevLett.95.260501} {\bibfield
  {journal} {\bibinfo  {journal} {Phys. Rev. Lett.}\ }\textbf {\bibinfo
  {volume} {95}},\ \bibinfo {pages} {260501} (\bibinfo {year}
  {2005}{\natexlab{a}})}\BibitemShut {NoStop}%
\bibitem [{\citenamefont {Ali-Khan}\ \emph {et~al.}(2007)\citenamefont
  {Ali-Khan}, \citenamefont {Broadbent},\ and\ \citenamefont
  {Howell}}]{Howell07}%
  \BibitemOpen
  \bibfield  {author} {\bibinfo {author} {\bibfnamefont {I.}~\bibnamefont
  {Ali-Khan}}, \bibinfo {author} {\bibfnamefont {C.~J.}\ \bibnamefont
  {Broadbent}}, \ and\ \bibinfo {author} {\bibfnamefont {J.~C.}\ \bibnamefont
  {Howell}},\ }\href {\doibase 10.1103/PhysRevLett.98.060503} {\bibfield
  {journal} {\bibinfo  {journal} {Phys. Rev. Lett.}\ }\textbf {\bibinfo
  {volume} {98}},\ \bibinfo {pages} {060503} (\bibinfo {year}
  {2007})}\BibitemShut {NoStop}%
\bibitem [{\citenamefont {Hillerkuss}\ \emph {et~al.}(2010)\citenamefont
  {Hillerkuss}, \citenamefont {Schellinger}, \citenamefont {Schmogrow},
  \citenamefont {Winter}, \citenamefont {Vallaitis}, \citenamefont {Bonk},
  \citenamefont {Marculescu}, \citenamefont {Li}, \citenamefont {Dreschmann},
  \citenamefont {Meyer}, \citenamefont {Ezra}, \citenamefont {Narkiss},
  \citenamefont {Nebendahl}, \citenamefont {Parmigiani}, \citenamefont
  {Petropoulos}, \citenamefont {Resan}, \citenamefont {Weingarten},
  \citenamefont {Ellermeyer}, \citenamefont {Lutz}, \citenamefont {M\"{o}ller},
  \citenamefont {H\"{u}bner}, \citenamefont {Becker}, \citenamefont {Koos},
  \citenamefont {Freude},\ and\ \citenamefont {Leuthold}}]{Hillerkuss10}%
  \BibitemOpen
  \bibfield  {author} {\bibinfo {author} {\bibfnamefont {D.}~\bibnamefont
  {Hillerkuss}}, \bibinfo {author} {\bibfnamefont {T.}~\bibnamefont
  {Schellinger}}, \bibinfo {author} {\bibfnamefont {R.}~\bibnamefont
  {Schmogrow}}, \bibinfo {author} {\bibfnamefont {M.}~\bibnamefont {Winter}},
  \bibinfo {author} {\bibfnamefont {T.}~\bibnamefont {Vallaitis}}, \bibinfo
  {author} {\bibfnamefont {R.}~\bibnamefont {Bonk}}, \bibinfo {author}
  {\bibfnamefont {A.}~\bibnamefont {Marculescu}}, \bibinfo {author}
  {\bibfnamefont {J.}~\bibnamefont {Li}}, \bibinfo {author} {\bibfnamefont
  {M.}~\bibnamefont {Dreschmann}}, \bibinfo {author} {\bibfnamefont
  {J.}~\bibnamefont {Meyer}}, \bibinfo {author} {\bibfnamefont {S.~B.}\
  \bibnamefont {Ezra}}, \bibinfo {author} {\bibfnamefont {N.}~\bibnamefont
  {Narkiss}}, \bibinfo {author} {\bibfnamefont {B.}~\bibnamefont {Nebendahl}},
  \bibinfo {author} {\bibfnamefont {F.}~\bibnamefont {Parmigiani}}, \bibinfo
  {author} {\bibfnamefont {P.}~\bibnamefont {Petropoulos}}, \bibinfo {author}
  {\bibfnamefont {B.}~\bibnamefont {Resan}}, \bibinfo {author} {\bibfnamefont
  {K.}~\bibnamefont {Weingarten}}, \bibinfo {author} {\bibfnamefont
  {T.}~\bibnamefont {Ellermeyer}}, \bibinfo {author} {\bibfnamefont
  {J.}~\bibnamefont {Lutz}}, \bibinfo {author} {\bibfnamefont {M.}~\bibnamefont
  {M\"{o}ller}}, \bibinfo {author} {\bibfnamefont {M.}~\bibnamefont
  {H\"{u}bner}}, \bibinfo {author} {\bibfnamefont {J.}~\bibnamefont {Becker}},
  \bibinfo {author} {\bibfnamefont {C.}~\bibnamefont {Koos}}, \bibinfo {author}
  {\bibfnamefont {W.}~\bibnamefont {Freude}}, \ and\ \bibinfo {author}
  {\bibfnamefont {J.}~\bibnamefont {Leuthold}},\ }in\ \href {\doibase
  10.1364/OFC.2010.PDPC1} {\emph {\bibinfo {booktitle} {Optical Fiber
  Communication Conference}}}\ (\bibinfo  {publisher} {Optical Society of
  America},\ \bibinfo {year} {2010})\ p.\ \bibinfo {pages} {PDPC1}\BibitemShut
  {NoStop}%
\bibitem [{\citenamefont {Hillerkuss}\ \emph {et~al.}(2011)\citenamefont
  {Hillerkuss}, \citenamefont {Schmogrow}, \citenamefont {Schellinger},
  \citenamefont {Jordan}, \citenamefont {Winter}, \citenamefont {Huber},
  \citenamefont {Vallaitis}, \citenamefont {Bonk}, \citenamefont {Kleinow},
  \citenamefont {Frey} \emph {et~al.}}]{Hillerkuss11}%
  \BibitemOpen
  \bibfield  {author} {\bibinfo {author} {\bibfnamefont {D.}~\bibnamefont
  {Hillerkuss}}, \bibinfo {author} {\bibfnamefont {R.}~\bibnamefont
  {Schmogrow}}, \bibinfo {author} {\bibfnamefont {T.}~\bibnamefont
  {Schellinger}}, \bibinfo {author} {\bibfnamefont {M.}~\bibnamefont {Jordan}},
  \bibinfo {author} {\bibfnamefont {M.}~\bibnamefont {Winter}}, \bibinfo
  {author} {\bibfnamefont {G.}~\bibnamefont {Huber}}, \bibinfo {author}
  {\bibfnamefont {T.}~\bibnamefont {Vallaitis}}, \bibinfo {author}
  {\bibfnamefont {R.}~\bibnamefont {Bonk}}, \bibinfo {author} {\bibfnamefont
  {P.}~\bibnamefont {Kleinow}}, \bibinfo {author} {\bibfnamefont
  {F.}~\bibnamefont {Frey}},  \emph {et~al.},\ }\href@noop {} {\bibfield
  {journal} {\bibinfo  {journal} {Nature Photonics}\ }\textbf {\bibinfo
  {volume} {5}},\ \bibinfo {pages} {364} (\bibinfo {year} {2011})}\BibitemShut
  {NoStop}%
\bibitem [{\citenamefont {Thuillier}\ and\ \citenamefont
  {Shepherd}(1985)}]{Thuillier85}%
  \BibitemOpen
  \bibfield  {author} {\bibinfo {author} {\bibfnamefont {G.}~\bibnamefont
  {Thuillier}}\ and\ \bibinfo {author} {\bibfnamefont {G.~G.}\ \bibnamefont
  {Shepherd}},\ }\href {\doibase 10.1364/AO.24.001599} {\bibfield  {journal}
  {\bibinfo  {journal} {Appl. Opt.}\ }\textbf {\bibinfo {volume} {24}},\
  \bibinfo {pages} {1599} (\bibinfo {year} {1985})}\BibitemShut {NoStop}%
\bibitem [{\citenamefont {Gault}\ \emph {et~al.}(1985)\citenamefont {Gault},
  \citenamefont {Johnston},\ and\ \citenamefont {Kendall}}]{Gault85}%
  \BibitemOpen
  \bibfield  {author} {\bibinfo {author} {\bibfnamefont {W.~A.}\ \bibnamefont
  {Gault}}, \bibinfo {author} {\bibfnamefont {S.~F.}\ \bibnamefont {Johnston}},
  \ and\ \bibinfo {author} {\bibfnamefont {D.~J.~W.}\ \bibnamefont {Kendall}},\
  }\href {\doibase 10.1364/AO.24.001604} {\bibfield  {journal} {\bibinfo
  {journal} {Appl. Opt.}\ }\textbf {\bibinfo {volume} {24}},\ \bibinfo {pages}
  {1604} (\bibinfo {year} {1985})}\BibitemShut {NoStop}%
\bibitem [{\citenamefont {Hsieh}(2009)}]{DLIPatent}%
  \BibitemOpen
  \bibfield  {author} {\bibinfo {author} {\bibfnamefont {Y.}~\bibnamefont
  {Hsieh}},\ }\href {http://www.google.com/patents/US7522343} {\enquote
  {\bibinfo {title} {Michelson interferometer based delay line
  interferometers},}\ } (\bibinfo {year} {2009}),\ \bibinfo {note} {uS Patent
  7,522,343}\BibitemShut {NoStop}%
\bibitem [{\citenamefont {Muller}\ \emph {et~al.}(1996)\citenamefont {Muller},
  \citenamefont {Zbinden},\ and\ \citenamefont {Gisin}}]{Muller96}%
  \BibitemOpen
  \bibfield  {author} {\bibinfo {author} {\bibfnamefont {A.}~\bibnamefont
  {Muller}}, \bibinfo {author} {\bibfnamefont {H.}~\bibnamefont {Zbinden}}, \
  and\ \bibinfo {author} {\bibfnamefont {N.}~\bibnamefont {Gisin}},\ }\href
  {http://stacks.iop.org/0295-5075/33/i=5/a=335} {\bibfield  {journal}
  {\bibinfo  {journal} {EPL (Europhysics Letters)}\ }\textbf {\bibinfo {volume}
  {33}},\ \bibinfo {pages} {335} (\bibinfo {year} {1996})}\BibitemShut
  {NoStop}%
\bibitem [{\citenamefont {Ribordy}\ \emph {et~al.}(1998)\citenamefont
  {Ribordy}, \citenamefont {Gautier}, \citenamefont {Gisin}, \citenamefont
  {Guinnard},\ and\ \citenamefont {Zbinden}}]{Ribordy98}%
  \BibitemOpen
  \bibfield  {author} {\bibinfo {author} {\bibfnamefont {G.}~\bibnamefont
  {Ribordy}}, \bibinfo {author} {\bibfnamefont {J.~D.}\ \bibnamefont
  {Gautier}}, \bibinfo {author} {\bibfnamefont {N.}~\bibnamefont {Gisin}},
  \bibinfo {author} {\bibfnamefont {O.}~\bibnamefont {Guinnard}}, \ and\
  \bibinfo {author} {\bibfnamefont {H.}~\bibnamefont {Zbinden}},\ }\href@noop
  {} {\bibfield  {journal} {\bibinfo  {journal} {Electronics Letters}\ }\textbf
  {\bibinfo {volume} {34}},\ \bibinfo {pages} {2116} (\bibinfo {year}
  {1998})}\BibitemShut {NoStop}%
\bibitem [{\citenamefont {Jain}\ \emph {et~al.}(2014)\citenamefont {Jain},
  \citenamefont {Anisimova}, \citenamefont {Khan}, \citenamefont {Makarov},
  \citenamefont {Marquardt},\ and\ \citenamefont {Leuchs}}]{Makarov2014}%
  \BibitemOpen
  \bibfield  {author} {\bibinfo {author} {\bibfnamefont {N.}~\bibnamefont
  {Jain}}, \bibinfo {author} {\bibfnamefont {E.}~\bibnamefont {Anisimova}},
  \bibinfo {author} {\bibfnamefont {I.}~\bibnamefont {Khan}}, \bibinfo {author}
  {\bibfnamefont {V.}~\bibnamefont {Makarov}}, \bibinfo {author} {\bibfnamefont
  {C.}~\bibnamefont {Marquardt}}, \ and\ \bibinfo {author} {\bibfnamefont
  {G.}~\bibnamefont {Leuchs}},\ }\href
  {http://stacks.iop.org/1367-2630/16/i=12/a=123030} {\bibfield  {journal}
  {\bibinfo  {journal} {New Journal of Physics}\ }\textbf {\bibinfo {volume}
  {16}},\ \bibinfo {pages} {123030} (\bibinfo {year} {2014})}\BibitemShut
  {NoStop}%
\bibitem [{\citenamefont {Gisin}\ \emph {et~al.}(2004)\citenamefont {Gisin},
  \citenamefont {Ribordy}, \citenamefont {Zbinden}, \citenamefont {Stucki},
  \citenamefont {Brunner},\ and\ \citenamefont {Scarani}}]{Valero04}%
  \BibitemOpen
  \bibfield  {author} {\bibinfo {author} {\bibfnamefont {N.}~\bibnamefont
  {Gisin}}, \bibinfo {author} {\bibfnamefont {G.}~\bibnamefont {Ribordy}},
  \bibinfo {author} {\bibfnamefont {H.}~\bibnamefont {Zbinden}}, \bibinfo
  {author} {\bibfnamefont {D.}~\bibnamefont {Stucki}}, \bibinfo {author}
  {\bibfnamefont {N.}~\bibnamefont {Brunner}}, \ and\ \bibinfo {author}
  {\bibfnamefont {V.}~\bibnamefont {Scarani}},\ }\href@noop {} {\enquote
  {\bibinfo {title} {Towards practical and fast quantum cryptography},}\ }
  (\bibinfo {year} {2004}),\ \Eprint
  {http://arxiv.org/abs/arXiv:quant-ph/0411022} {arXiv:quant-ph/0411022}
  \BibitemShut {NoStop}%
\bibitem [{\citenamefont {Moroder}\ \emph {et~al.}(2012)\citenamefont
  {Moroder}, \citenamefont {Curty}, \citenamefont {Lim}, \citenamefont {Thinh},
  \citenamefont {Zbinden},\ and\ \citenamefont {Gisin}}]{Tobias12}%
  \BibitemOpen
  \bibfield  {author} {\bibinfo {author} {\bibfnamefont {T.}~\bibnamefont
  {Moroder}}, \bibinfo {author} {\bibfnamefont {M.}~\bibnamefont {Curty}},
  \bibinfo {author} {\bibfnamefont {C.~C.~W.}\ \bibnamefont {Lim}}, \bibinfo
  {author} {\bibfnamefont {L.~P.}\ \bibnamefont {Thinh}}, \bibinfo {author}
  {\bibfnamefont {H.}~\bibnamefont {Zbinden}}, \ and\ \bibinfo {author}
  {\bibfnamefont {N.}~\bibnamefont {Gisin}},\ }\href {\doibase
  10.1103/PhysRevLett.109.260501} {\bibfield  {journal} {\bibinfo  {journal}
  {Phys. Rev. Lett.}\ }\textbf {\bibinfo {volume} {109}},\ \bibinfo {pages}
  {260501} (\bibinfo {year} {2012})}\BibitemShut {NoStop}%
\bibitem [{\citenamefont {Inoue}\ \emph {et~al.}(2002)\citenamefont {Inoue},
  \citenamefont {Waks},\ and\ \citenamefont {Yamamoto}}]{Yamamoto02}%
  \BibitemOpen
  \bibfield  {author} {\bibinfo {author} {\bibfnamefont {K.}~\bibnamefont
  {Inoue}}, \bibinfo {author} {\bibfnamefont {E.}~\bibnamefont {Waks}}, \ and\
  \bibinfo {author} {\bibfnamefont {Y.}~\bibnamefont {Yamamoto}},\ }\href
  {\doibase 10.1103/PhysRevLett.89.037902} {\bibfield  {journal} {\bibinfo
  {journal} {Phys. Rev. Lett.}\ }\textbf {\bibinfo {volume} {89}},\ \bibinfo
  {pages} {037902} (\bibinfo {year} {2002})}\BibitemShut {NoStop}%
\bibitem [{\citenamefont {Sasaki}\ \emph {et~al.}(2014)\citenamefont {Sasaki},
  \citenamefont {Yamamoto},\ and\ \citenamefont {Koashi}}]{Koashi14}%
  \BibitemOpen
  \bibfield  {author} {\bibinfo {author} {\bibfnamefont {T.}~\bibnamefont
  {Sasaki}}, \bibinfo {author} {\bibfnamefont {Y.}~\bibnamefont {Yamamoto}}, \
  and\ \bibinfo {author} {\bibfnamefont {M.}~\bibnamefont {Koashi}},\ }\href
  {\doibase 10.1038/nature13303} {\bibfield  {journal} {\bibinfo  {journal}
  {Nature}\ }\textbf {\bibinfo {volume} {509}},\ \bibinfo {pages} {475}
  (\bibinfo {year} {2014})}\BibitemShut {NoStop}%
\bibitem [{\citenamefont {Guan}\ \emph {et~al.}(2015)\citenamefont {Guan},
  \citenamefont {Cao}, \citenamefont {Liu}, \citenamefont {Shen-Tu},
  \citenamefont {Pelc}, \citenamefont {Fejer}, \citenamefont {Peng},
  \citenamefont {Ma}, \citenamefont {Zhang},\ and\ \citenamefont
  {Pan}}]{Ma15RR}%
  \BibitemOpen
  \bibfield  {author} {\bibinfo {author} {\bibfnamefont {J.-Y.}\ \bibnamefont
  {Guan}}, \bibinfo {author} {\bibfnamefont {Z.}~\bibnamefont {Cao}}, \bibinfo
  {author} {\bibfnamefont {Y.}~\bibnamefont {Liu}}, \bibinfo {author}
  {\bibfnamefont {G.-L.}\ \bibnamefont {Shen-Tu}}, \bibinfo {author}
  {\bibfnamefont {J.~S.}\ \bibnamefont {Pelc}}, \bibinfo {author}
  {\bibfnamefont {M.~M.}\ \bibnamefont {Fejer}}, \bibinfo {author}
  {\bibfnamefont {C.-Z.}\ \bibnamefont {Peng}}, \bibinfo {author}
  {\bibfnamefont {X.}~\bibnamefont {Ma}}, \bibinfo {author} {\bibfnamefont
  {Q.}~\bibnamefont {Zhang}}, \ and\ \bibinfo {author} {\bibfnamefont {J.-W.}\
  \bibnamefont {Pan}},\ }\href {\doibase 10.1103/PhysRevLett.114.180502}
  {\bibfield  {journal} {\bibinfo  {journal} {Phys. Rev. Lett.}\ }\textbf
  {\bibinfo {volume} {114}},\ \bibinfo {pages} {180502} (\bibinfo {year}
  {2015})}\BibitemShut {NoStop}%
\bibitem [{\citenamefont {Takesue}\ \emph {et~al.}(2015)\citenamefont
  {Takesue}, \citenamefont {Sasaki}, \citenamefont {Tamaki},\ and\
  \citenamefont {Koashi}}]{Koashi15}%
  \BibitemOpen
  \bibfield  {author} {\bibinfo {author} {\bibfnamefont {H.}~\bibnamefont
  {Takesue}}, \bibinfo {author} {\bibfnamefont {T.}~\bibnamefont {Sasaki}},
  \bibinfo {author} {\bibfnamefont {K.}~\bibnamefont {Tamaki}}, \ and\ \bibinfo
  {author} {\bibfnamefont {M.}~\bibnamefont {Koashi}},\ }\href@noop {}
  {\bibfield  {journal} {\bibinfo  {journal} {Nature Photonics}\ } (\bibinfo
  {year} {2015})}\BibitemShut {NoStop}%
\bibitem [{\citenamefont {Franson}(1989)}]{Franson89}%
  \BibitemOpen
  \bibfield  {author} {\bibinfo {author} {\bibfnamefont {J.~D.}\ \bibnamefont
  {Franson}},\ }\href {\doibase 10.1103/PhysRevLett.62.2205} {\bibfield
  {journal} {\bibinfo  {journal} {Phys. Rev. Lett.}\ }\textbf {\bibinfo
  {volume} {62}},\ \bibinfo {pages} {2205} (\bibinfo {year}
  {1989})}\BibitemShut {NoStop}%
\bibitem [{\citenamefont {Dada}\ \emph {et~al.}(2011)\citenamefont {Dada},
  \citenamefont {Leach}, \citenamefont {Buller}, \citenamefont {Padgett},\ and\
  \citenamefont {Andersson}}]{Erika11}%
  \BibitemOpen
  \bibfield  {author} {\bibinfo {author} {\bibfnamefont {A.~C.}\ \bibnamefont
  {Dada}}, \bibinfo {author} {\bibfnamefont {J.}~\bibnamefont {Leach}},
  \bibinfo {author} {\bibfnamefont {G.~S.}\ \bibnamefont {Buller}}, \bibinfo
  {author} {\bibfnamefont {M.~J.}\ \bibnamefont {Padgett}}, \ and\ \bibinfo
  {author} {\bibfnamefont {E.}~\bibnamefont {Andersson}},\ }\href@noop {}
  {\bibfield  {journal} {\bibinfo  {journal} {Nature Physics}\ }\textbf
  {\bibinfo {volume} {7}},\ \bibinfo {pages} {677} (\bibinfo {year}
  {2011})}\BibitemShut {NoStop}%
\bibitem [{\citenamefont {Barreiro}\ \emph
  {et~al.}(2005{\natexlab{b}})\citenamefont {Barreiro}, \citenamefont
  {Langford}, \citenamefont {Peters},\ and\ \citenamefont {Kwiat}}]{Kwiat97}%
  \BibitemOpen
  \bibfield  {author} {\bibinfo {author} {\bibfnamefont {J.~T.}\ \bibnamefont
  {Barreiro}}, \bibinfo {author} {\bibfnamefont {N.~K.}\ \bibnamefont
  {Langford}}, \bibinfo {author} {\bibfnamefont {N.~A.}\ \bibnamefont
  {Peters}}, \ and\ \bibinfo {author} {\bibfnamefont {P.~G.}\ \bibnamefont
  {Kwiat}},\ }\href@noop {} {\bibfield  {journal} {\bibinfo  {journal}
  {Physical Review Letters}\ }\textbf {\bibinfo {volume} {95}},\ \bibinfo
  {pages} {260501} (\bibinfo {year} {2005}{\natexlab{b}})}\BibitemShut
  {NoStop}%
\bibitem [{\citenamefont {Lo}\ \emph {et~al.}(2005)\citenamefont {Lo},
  \citenamefont {Ma},\ and\ \citenamefont {Chen}}]{Lodecoy05}%
  \BibitemOpen
  \bibfield  {author} {\bibinfo {author} {\bibfnamefont {H.-K.}\ \bibnamefont
  {Lo}}, \bibinfo {author} {\bibfnamefont {X.}~\bibnamefont {Ma}}, \ and\
  \bibinfo {author} {\bibfnamefont {K.}~\bibnamefont {Chen}},\ }\href {\doibase
  10.1103/PhysRevLett.94.230504} {\bibfield  {journal} {\bibinfo  {journal}
  {Phys. Rev. Lett.}\ }\textbf {\bibinfo {volume} {94}},\ \bibinfo {pages}
  {230504} (\bibinfo {year} {2005})}\BibitemShut {NoStop}%
\bibitem [{\citenamefont {Fulop}(2015)}]{Ludvoic}%
  \BibitemOpen
  \bibfield  {author} {\bibinfo {author} {\bibfnamefont {L.}~\bibnamefont
  {Fulop}},\ }\href@noop {} {}\bibinfo {howpublished} {Kylia, 10 Rue de
  Montmorency, 75003 Paris, France} (\bibinfo {year} {personal communication,
  2015})\BibitemShut {NoStop}%
\end{thebibliography}


%

\end{document}